\documentclass[journal]{IEEEtran}

\usepackage{cite}
\usepackage{amsmath,amssymb,amsfonts}
\usepackage{algorithm}
\usepackage{algpseudocode}
\usepackage{textcomp}
\usepackage{xcolor}
\usepackage{booktabs}
\usepackage{array}
\usepackage{siunitx}
\usepackage{subcaption}
\usepackage{graphicx}
\usepackage{dblfloatfix}
\usepackage{placeins}
\usepackage{cuted}
\usepackage[utf8]{inputenc}
\usepackage[T1]{fontenc}
\usepackage{enumitem}
\usepackage{mathtools}

\usepackage[
  hidelinks,
  linktoc=all,
  bookmarksnumbered=true,
  bookmarksopen=true,
  breaklinks=true
]{hyperref}

\hypersetup{
  pdftitle={Low-Complexity Run-Length-Limited ISI-Mitigation (RLIM) Codes for Molecular Communication},
  pdfauthor={Melih Sahin and Ozgur B. Akan},
  pdfsubject={Molecular communication, constrained coding, enumerative coding, RLIM coding},
  pdfkeywords={Constrained coding, enumerative coding, molecular communication, RLIM coding, run-length-limited coding}
}

\hypersetup{
    colorlinks=true,
    linkcolor=blue,
    citecolor=blue,
    urlcolor=blue
}

\usepackage{orcidlink}

\title{Low-Complexity Run-Length-Limited ISI-Mitigation (RLIM) Codes for Molecular Communication}

\author{Melih~\c{S}ahin \orcidlink{0009-0000-6813-3297},~\IEEEmembership{Graduate Student Member,~IEEE,}
Ozgur~B.~Akan \orcidlink{0000-0003-2523-3858},~\IEEEmembership{Fellow,~IEEE}
\thanks{Melih \c{S}ahin is with the Centre for neXt Communications (CXC), Department of Engineering, University of Cambridge, CB3 0FA  Cambridge, U.K. (e-mail: \href{mailto:ms3195@cam.ac.uk}{ms3195@cam.ac.uk}).}
\thanks{Ozgur B. Akan is with the Centre for neXt Communications (CXC), Department of Engineering, University of Cambridge, CB3 0FA Cambridge, U.K., and also with the Centre for neXt Communications (CXC), Department of Electrical and Electronics Engineering, Ko\c{c} University, 34450 Istanbul, T\"urkiye (e-mail: \href{mailto:oba21@cam.ac.uk}{oba21@cam.ac.uk}).}
\thanks{This work was supported in part by the AXA Research Fund (AXA Chair for Internet of Everything at Ko\c{c} University).}
}



\begin{document}
\maketitle

\begin{abstract}
Molecular communication suffers from severe inter-symbol interference, which makes constrained coding essential for reliable transmission. Run-length-limited ISI-mitigation codes are attractive because they select low-weight constrained codebooks, reducing ISI while allowing more molecules to be assigned to each transmitted \(1\)-symbol under the usual molecular-communication normalization. Previous results showed strong bit-error-rate performance for these codes, but their original realization required full codebook generation and storage. This exponential storage growth is unsuitable for resource-constrained molecular communication channels and also limits the exploration of larger information dimensions. This is particularly important for nano-scale molecular communication, where transmitter and receiver nodes are expected to operate under severe memory and computational constraints. This paper removes that realization bottleneck by replacing full codebook storage with an enumerative realization based on Cover's ranking framework, constant-weight run-length-limited counting, and cumulative weight-layer offsets. The resulting encoder and decoder preserve the selected RLIM codebooks and the original projection-based decoding behavior while storing only polynomial-size counting tables. Storage and runtime measurements confirm the resulting exponential-to-polynomial reduction, while diffusion-based molecular-communication simulations show that the newly accessible larger information dimensions can improve the best mean bit-error-rate performance attainable by individual RLIM orders in several tested channel regimes.

\end{abstract}


\section{Introduction}

Molecular communication (MC) conveys information through chemical signals rather than electromagnetic waves \cite{Nakano2013,Akan2017Fund}. It is a natural communication paradigm for environments where electromagnetic channels are inefficient, strongly attenuated, or physically unsuitable, including in-body nanonetworks, targeted drug delivery, lab-on-chip systems, synthetic biological systems, and bio-hybrid sensing networks \cite{Akyildiz2010,Kuscu2019}. In diffusion-based MC, however, molecules released in one signaling interval may continue arriving in many later intervals, producing severe inter-symbol interference (ISI) \cite{Nakano2013}. Since the reliability of a received symbol depends strongly on the surrounding transmitted pattern, constrained-sequence design is especially natural in this setting.

Run-length-limited ISI-mitigation (RLIM) coding was introduced in \cite{SahinAkan2024RLIM} for diffusion-based MC. RLIM selects a size-\(2^k\) subset with minimum total Hamming weight from an admissible run-length-limited family. Under the usual MC normalization, this is useful because low-weight codebooks reduce the number of transmitted \(1\)-symbols, allowing more molecules to be assigned to each transmitted pulse, while the run-length constraint suppresses ISI. In the tested diffusion-based MC settings of \cite{SahinAkan2024RLIM}, RLIM was the strongest coding scheme when compared with classical RLL codes and other prominent MC coding baselines.

The main obstacle to using RLIM in resource-constrained MC devices is its lookup-table realization. In that realization, the admissible constrained family is generated, the minimum-weight subset is extracted, and the selected codebook is stored in full. Hence the storage grows exponentially with the number of information bits per block. This is unsuitable for nanoscale or highly resource-limited MC transmitters and receivers, and it also limits the exploration of larger information dimensions.

This paper removes the full-codebook storage bottleneck by realizing RLIM encoding and decoding as ranking and unranking operations. We combine Cover's enumerative framework \cite{Cover1973} with classical constant-weight run-length-limited counting \cite{Kurmaev2002,Kurmaev2009}. The RLIM-specific step is the use of cumulative lower-weight offsets, which recover the original minimum-weight RLIM ordering across Hamming-weight layers. Thus the same selected codebooks and the same projection-based decoding convention of RLIM codes are preserved, but the selected codewords no longer need to be stored.

The storage reduction is exponential-to-polynomial. For example, for order \(i=3\) and information dimension \(k=16\), which is one of the main regimes tested in the original RLIM study \cite{SahinAkan2024RLIM}, full codebook storage requires \(2{,}424{,}832\) bits, whereas the proposed counting tables require only \(9{,}359\) bits in our implementation. Thus, even at this moderate dimension, the stored representation is reduced by about \(259\times\), and the gap becomes much larger as \(k\) increases. This reduction takes one of the strongest tested MC channel-coding methods from a storage-limited lookup-table construction toward an implementable coding mechanism for future resource-constrained micro- and nanoscale MC platforms.

Beyond storage, codeword recovery becomes a single ranking scan rather than a search over a stored codebook, and projection decoding removes the multiplicative dependence on the information dimension that appears in the stored-codebook implementation. The proposed realization also makes larger selected RLIM codebooks directly testable in a simulation setting. This is important because, among the prominent MC coding schemes compared in this paper, moderate-order RLIM codes attain the strongest overall BER performance across the tested operating points, and the reported \(k\)-sweeps show that information dimensions beyond the \(k=16\) regime considered with the original full-codebook realization can improve the BER performance of individual RLIM orders.

The remainder of this paper is organized as follows. Section II gives the MC channel model used in this paper. Section III defines the RLIM family and the selected codebooks. Section IV develops the ranking framework. Section V gives the algorithms and complexity analysis. Section VI gives the simulation framework and numerical results. Section VII concludes the paper.

\section{Molecular Communication Model}
\label{sec:model}

\begin{figure}[t]
\centering
\includegraphics[width=0.47\textwidth]{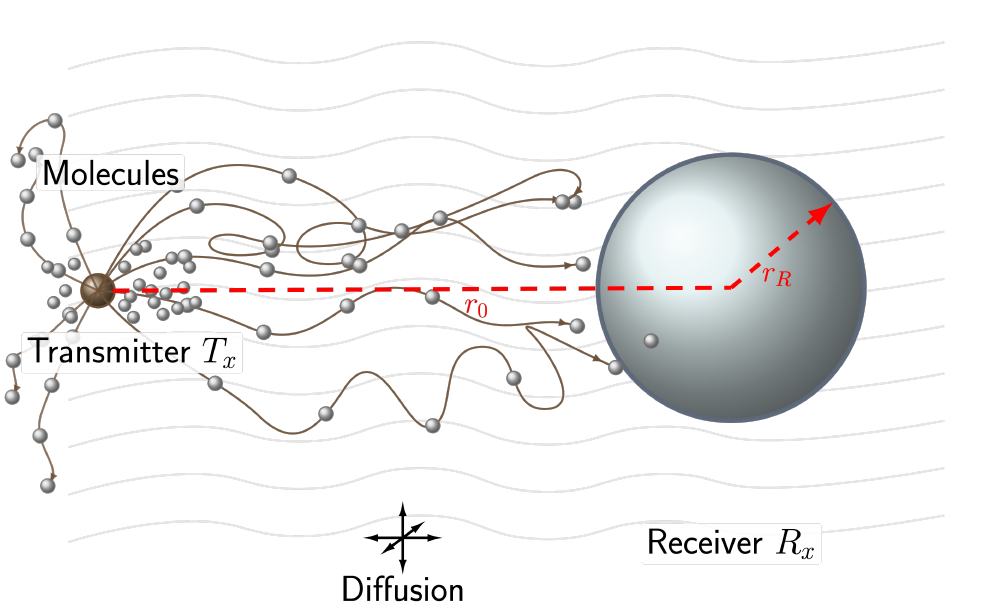}
\caption{MC Channel Model}
\label{fig:channel}
\end{figure}
We adopt the standard single-input single-output diffusion-based MC channel with binary concentration-shift keying modulation \cite{BCSK} and a fully absorbing spherical receiver \cite{Nakano2013}, a widely used model for in-body Nano-IoT channels \cite{Kuscu2019}. The corresponding channel geometry is depicted in Fig. \ref{fig:channel} \cite{ldp}. For this absorbing-receiver geometry, the cumulative probability that a molecule emitted by the transmitter has been absorbed by time \(t\) is
\begin{equation}
F(t)=\frac{r_R}{r_0}\cdot
\mathrm{erfc}\!\left(
\frac{r_0-r_R}{\sqrt{4Dt}}
\right),
\label{eq:mc_F}
\end{equation}
where \(D\) is the diffusion coefficient, \(r_R\) is the receiver radius, and \(r_0\) is the distance from the transmitter to the centre of the receiver \cite{channel_characteristics}.

For a signal interval duration \(t_{s}\), the probability that a molecule released at the beginning of an interval is absorbed during the \(\ell\)-th subsequent interval is \cite{Kuscu2019}
\begin{equation}
p_\ell = F(\ell \cdot t_{s})-F((\ell-1) \cdot t_{s}),
\qquad
\ell\ge 1.
\label{eq:mc_taps}
\end{equation}
To obtain a finite-memory simulation model with memory \(I\), the probability mass outside the retained channel taps is written as
\begin{equation}
\pi_{\mathrm{tail}}
=
1-\sum_{\ell=1}^{I}p_\ell.
\label{eq:mc_tail}
\end{equation}
Then, via \cite{multinomial}, a single emission of \(M\) molecules is represented by
\begin{equation}
(X_1,\dots,X_I,X_{I+1})
\sim
\mathrm{Multinomial}\!\bigl(
M;\, p_1,\dots,p_I,\pi_{\mathrm{tail}}
\bigr),
\label{eq:mc_multinomial}
\end{equation}
where \(X_\ell\) is the number of molecules absorbed \(\ell\) intervals after the emission, and \(X_{I+1}\) accounts for molecules absorbed after the simulated memory window, or not absorbed by the receiver.

For a transmitted bit stream \(b_t\in\{0,1\}\), the received count in interval \(t\) is modeled as
\begin{equation}
N_t
=
\sum_{\ell=1}^{\min\{t,I\}}
b_{t-\ell+1}\cdot X_{\ell}^{(t-\ell+1)}
+
W_t, 
\label{eq:mc_received}
\end{equation}
where \(X_{\ell}^{(t-\ell+1)}\) denotes the contribution of the emission made in interval \(t-\ell+1\) and absorbed \(\ell\) intervals later, and \(W_t\sim \lfloor \mathcal{N}(0,\sigma^2)\rceil\) models receiver counting noise \cite{ldp}.

\section{Run-Length-Limited ISI-Mitigation (RLIM) Codes}
\label{sec:rlim_family}

This section recalls the RLIM code construction from \cite{SahinAkan2024RLIM}. First, we fix the RLIM order \(i\in\mathbb{N}\) and the code length \(n\ge i+1\). For \(a\in\{0,1\}\) and \(m\in\mathbb{N}_0\), let \(a^m\) denote the binary word formed by repeating the symbol \(a\) \(m\) times. For binary words \(\alpha\) and \(\beta\), let \(\alpha\circ\beta\) denote their concatenation. Let \(\varepsilon\) denote the empty word.

For \(t\in\mathbb{N}_0\), let \(\mathcal{C}_i(t)\subseteq\{0,1\}^t\) denote the set of binary \((i,\infty)\)-RLL sequences of length \(t\), that is, the set of binary words of length \(t\) in which any two successive \(1\)-bits are separated by at least \(i\) zeros \cite{BeenkerImmink1983,HareedyDabakCalderbank2022}. The base cases are
\begin{equation}
\mathcal{C}_i(0)=\{\varepsilon\},
\end{equation}
and, for \(1\le t\le i+1\),
\begin{equation}
\mathcal{C}_i(t)
=
\{\,0^t\,\}
\cup
\{\,0^{\,t-r}\circ 1\circ 0^{\,r-1}: r=1,2,\dots,t\,\}.
\label{eq:base_family}
\end{equation}
For \(t\ge i+2\), any word in \(\mathcal{C}_i(t)\) begins in one of two ways. Either it begins with \(0\), followed by a word in \(\mathcal{C}_i(t-1)\), or it begins with \(1\circ 0^i\), followed by a word in \(\mathcal{C}_i(t-i-1)\). Hence, following \cite{recursion},
\begin{equation}
\mathcal{C}_i(t)
=
\bigl\{\,0\circ \mathbf{u}:\mathbf{u}\in\mathcal{C}_i(t-1)\,\bigr\}
\cup
\bigl\{\,1\circ 0^i\circ \mathbf{u}:\mathbf{u}\in\mathcal{C}_i(t-i-1)\,\bigr\}.
\label{eq:Ci_recursion}
\end{equation}
This recursive generation order is lexicographic.

In this paper, we use the enhanced RLIM family of length \(n\), which includes the all-zero word, and denote it simply by \(\mathrm{RLIM}_i(n)\):
\begin{equation}
\mathrm{RLIM}_i(n)
=
\bigl\{\,0^i\circ \mathbf{u}:\mathbf{u}\in\mathcal{C}_i(n-i)\,\bigr\}.
\label{eq:RLIM_full}
\end{equation}
Thus, throughout the paper, the notation \(\mathrm{RLIM}_i(n)\) refers to the enhanced RLIM family unless explicitly stated otherwise. The original non-enhanced RLIM realization \cite{SahinAkan2024RLIM} is obtained by excluding the all-zero word \(0^n\) from \(\mathrm{RLIM}_i(n)\). This distinction originates from the detection method. In the original adaptive/dynamic-threshold setting, each transmitted codeword is required to contain at least one \(1\)-bit, and therefore the all-zero word is excluded. In contrast, under the static-threshold detection considered in this paper for non-drift channels, this restriction is unnecessary, so the enhanced family including \(0^n\) is used.

Accordingly, all ranking, unranking, encoding, and decoding developments in the remainder of the paper are presented for the enhanced family \(\mathrm{RLIM}_i(n)\). The same realization applies to the non-enhanced case with only a minor change in the selected rank range, as explained later in Section~\ref{sec:algorithms}.

For each length \(n\), let
\begin{equation}
\mathcal O_i(n)=\bigl(x_n^{(0)},x_n^{(1)},x_n^{(2)},\dots\bigr)
\label{eq:ordered_family_hat}
\end{equation}
denote the words in \(\mathrm{RLIM}_i(n)\), ordered first by Hamming weight and then, within each fixed weight, lexicographically.

The selected RLIM codebook, when ordered, is
\begin{equation}
\mathrm{RLIM}_i(n,k)
=
(x_n^{(0)},x_n^{(1)},\dots,x_n^{(2^k-1)}),
\label{eq:enhanced_codebook}
\end{equation}
where \(n\) is the selected codeword length. If one takes the shortest admissible length, then
\begin{equation}
n
=
\min\{\,\nu\ge i+1:\ |\mathcal{C}_i(\nu-i)|\ge 2^k\,\}.
\label{eq:n1_def}
\end{equation}

\section{Efficient Realization of RLIM Codes}
\label{sec:cover}

We begin with Cover's classical rank formula in its general form \cite{Cover1973}. Let \(L\in\mathbb{N}\), and let
\begin{equation}
S\subseteq\{0,1\}^{L}
\end{equation}
be any binary code of fixed length \(L\), ordered lexicographically. For a prefix \((a_1,\dots,a_j)\), let
\begin{equation}
n_S(a_1,\dots,a_j)
\end{equation}
denote the number of codewords in \(S\) whose first \(j\) bits are \((a_1,\dots,a_j)\). Then, for every
\begin{equation}
\mathbf{x}=(x_1,\dots,x_L)\in S,
\end{equation}
the corresponding lexicographic rank in \(S\) is
\begin{equation}
\operatorname{rank}_{S}(\mathbf{x})
=
\sum_{j=1}^{L} x_j\cdot n_S(x_1,\dots,x_{j-1},0).
\label{eq:cover_general}
\end{equation}

The inverse unranking procedure is also straightforward \cite{Cover1973}. Given a target rank \(r\in\{0,1,\dots,|S|-1\}\), suppose \(x_1,\dots,x_{j-1}\) has already been fixed, and define
\begin{equation}
c_j=n_S(x_1,\dots,x_{j-1},0).
\label{eq:general_unrank_count}
\end{equation}
Then, at step \(j\),
\begin{equation}
x_j=
\begin{cases}
0, & r<c_j,\\
1, & r\ge c_j,
\end{cases}
\qquad
r\leftarrow
\begin{cases}
r, & r<c_j,\\
r-c_j, & r\ge c_j.
\end{cases}
\label{eq:general_unranking_rule}
\end{equation}
Repeating \eqref{eq:general_unranking_rule} for \(j=1,\dots,L\) yields the unique codeword of the prescribed rank.

We now specialize this framework to the RLIM family. Fix a codeword length \(n\), and set
\begin{equation}
T=n-i.
\label{eq:Tdef}
\end{equation}
Every word in \(\mathrm{RLIM}_i(n)\) has the form
\begin{equation}
0^i\circ \mathbf{u},
\end{equation}
where \(\mathbf{u}\in\{0,1\}^{T}\). Thus, once the codeword length \(n\) is fixed, ranking and unranking reduce to subsets of \(\{0,1\}^{T}\). The counting formulas below are therefore developed directly for \(\mathrm{RLIM}_i(n)\).

For \(w\in\{0,1,\dots,T\}\), define the fixed-weight subcode
\begin{equation}
\mathcal{C}_i(T,w)=\{\mathbf{u}\in\mathcal{C}_i(T): \mathrm{weight}(\mathbf{u})=w\}.
\label{eq:fixed_weight_subcode}
\end{equation}
For a binary prefix \(q\) of length at most \(T\), define
\begin{equation}
n_{\mathcal{C}_i(T,w)}(q)
=
\bigl|\{\mathbf{u}\in\mathcal{C}_i(T,w):\mathbf{u}\text{ begins with }q\}\bigr|.
\label{eq:fixed_weight_prefix_count}
\end{equation}
Then Cover's formula Eq. \ref{eq:cover_general} gives the fixed-weight rank
\begin{equation}
\operatorname{rank}_{\mathcal{C}_i(T,w)}(\mathbf{u})
=
\sum_{j=1}^{T} u_j\cdot
n_{\mathcal{C}_i(T,w)}(u_1,\dots,u_{j-1},0).
\label{eq:cover_fixed_weight}
\end{equation}

For \(\ell\in\mathbb{N}_0\), \(s\in\mathbb{Z}\), and \(r\in\{0,1,\dots,i\}\), define
\begin{equation}
\mathcal{F}_i(\ell,s,r)
=
\left\{
\begin{array}{l}
\mathbf{v}\in\mathcal{C}_i(\ell):
\ \mathrm{weight}(\mathbf{v})=s,\\[1mm]
v_1=\cdots=v_{\min\{r,\ell\}}=0
\end{array}
\right\},
\label{eq:Fset}
\end{equation}
and let
\begin{equation}
F_i(\ell,s,r)=\bigl|\mathcal{F}_i(\ell,s,r)\bigr|.
\label{eq:Fdef}
\end{equation}
Thus \(F_i(\ell,s,r)\) counts the length-\(\ell\) \((i,\infty)\)-RLL words of weight \(s\) whose first \(\min\{r,\ell\}\) positions are fixed to zero. These quantities satisfy

\begin{flalign}
&F_i(0,0,r)=1, \qquad F_i(0,s,r)=0 \ \text{for } s\neq 0, && \label{eq:Fbase}\\
&F_i(\ell,s,r)=0 \ \text{for } \ell<0 \text{ or } s<0, && \label{eq:Finvalid}\\
&F_i(\ell,s,r)=F_i(\ell-1,s,r-1) \ \text{for } \ell\ge 1 \text{ and } r\ge 1, && \label{eq:Fforced}\\
&F_i(\ell,s,0)=F_i(\ell-1,s,0)+F_i(\ell-1,s-1,i) \ \text{for } \ell\ge 1 && \label{eq:Ffree}
\end{flalign}

Indeed, if \(r\ge 1\), the first bit is forced to be \(0\), so deleting it gives a word in \(\mathcal{F}_i(\ell-1,s,r-1)\). If \(r=0\), the first bit can be either \(0\) or \(1\). In the \(0\)-case, deleting it gives a word in \(\mathcal{F}_i(\ell-1,s,0)\). In the \(1\)-case, the remaining suffix has length \(\ell-1\), weight \(s-1\), and must begin with \(i\) zeros, so it lies in \(\mathcal{F}_i(\ell-1,s-1,i)\). This is the standard constant-weight run-length-limited enumeration idea \cite{Kurmaev2002,Kurmaev2009}, written here in the compact state variables \((\ell,s,r)\).

Now let \(q=q_1q_2\dots q_m\) be a binary prefix of length \(m\le T\). We call \(q\) valid if any two successive \(1\)-bits in \(q\) are separated by at least \(i\) zeros. For a valid prefix \(q\), define
\begin{equation}
h(q)=\sum_{j=1}^{m} q_j
\label{eq:hq}
\end{equation}
and
\begin{equation}
r(q)=
\begin{cases}
0, & \text{if } q \text{ contains no \(1\)-bit},\\[1mm]
\max\{0,\;i-z(q)\}, & \text{otherwise},
\end{cases}
\label{eq:rq}
\end{equation}
where \(z(q)\) is the number of trailing zeros after the last \(1\) in \(q\). In the prefix-count setting of constant-weight RLL enumeration \cite{Kurmaev2002}, a valid prefix \(q\) leaves a suffix of length \(T-m\), remaining weight \(w-h(q)\), and \(r(q)\) leading zeros that the suffix must contain. Therefore, the prefix count can be written as
\begin{equation}
n_{\mathcal{C}_i(T,w)}(q)=
\begin{cases}
F_i(T-m,\;w-h(q),\;r(q)), & q \text{ valid},\\[1mm]
0, & q \text{ invalid}.
\end{cases}
\label{eq:Bformula}
\end{equation}

Now let \(\mathbf{u}=(u_1,\dots,u_T)\in\mathcal{C}_i(T,w)\). For each \(j=1,\dots,T\), define
\begin{equation}
q_j=(u_1,\dots,u_{j-1},0).
\label{eq:qj_def}
\end{equation}
Then \eqref{eq:cover_fixed_weight} and \eqref{eq:Bformula} give
\begin{equation}
\operatorname{rank}_{\mathcal{C}_i(T,w)}(\mathbf{u})
=
\sum_{j=1}^{T}
u_j\cdot
F_i\!\bigl(T-j,\;w-h(q_j),\;r(q_j)\bigr).
\label{eq:within_subcode_rank}
\end{equation}

To pass from the rank inside a fixed-weight subcode to the full RLIM order, we also need the cardinality of each fixed-weight subcode. Since \(F_i(T,w,0)\) counts all admissible length-\(T\) words of weight \(w\) with no additional leading-zero constraint, we have
\begin{equation}
|\mathcal{C}_i(T,w)|=F_i(T,w,0).
\label{eq:NfromB}
\end{equation}
Hence the number of admissible internal words whose weight is strictly smaller than \(w\) is
\begin{equation}
\Gamma_i(T,w)=\sum_{a=0}^{w-1}F_i(T,a,0),
\label{eq:cumulative_layers}
\end{equation}
with \(\Gamma_i(T,0)=0\), and
\(\Gamma_i(T,T+1)=\sum_{a=0}^{T}F_i(T,a,0)=|\mathcal{C}_i(T)|\).

Therefore, if \(\mathbf{x}=0^i\circ\mathbf{u}\in\mathrm{RLIM}_i(n)\) and \(\mathbf{u}\in\mathcal{C}_i(T,w)\), then the rank of \(\mathbf{x}\) in the RLIM order is
\begin{equation}
\operatorname{rank}_{\mathrm{RLIM}_i(n)}(\mathbf{x})
=
\Gamma_i(T,w)
+
\operatorname{rank}_{\mathcal{C}_i(T,w)}(\mathbf{u}).
\label{eq:global_rank}
\end{equation}

Equation~\eqref{eq:global_rank} is the step that turns constant-weight RLL enumeration into an RLIM realization. The cumulative offset \(\Gamma_i(T,w)\) places each fixed-weight rank in the global minimum-weight RLIM order, so ranking and unranking reproduce the selected RLIM codebook without storing its codewords.

As a small example, let \(i=1\), \(n=4\), and \(k=2\), so \(T=3\). Then \(\mathcal{C}_1(3)=\{000,001,010,100,101\}\), and the selected RLIM codebook is \((0000,0001,0010,0100)\). For \(\mathbf{x}=0100\), the internal word is \(\mathbf{u}=100\) with weight \(w=1\). Since \(\Gamma_1(3,1)=1\) and \(\operatorname{rank}_{\mathcal{C}_1(3,1)}(100)=2\), Eq.~\eqref{eq:global_rank} gives global rank \(3\), matching its position in the selected RLIM order.

This process also gives the unranking procedure. Given a target rank \(r\), we first find the unique \(w^\star\) such that
\begin{equation}
\Gamma_i(T,w^\star)\le r<\Gamma_i(T,w^\star+1),
\label{eq:layer_choice}
\end{equation}
then set
\begin{equation}
\rho=r-\Gamma_i(T,w^\star),
\label{eq:within_rank}
\end{equation}
and reconstruct the unique word in the fixed-weight subcode \(\mathcal{C}_i(T,w^\star)\) having within-subcode rank \(\rho\) by the general unranking rule \eqref{eq:general_unranking_rule}.

\section{Algorithms and Complexity}
\label{sec:algorithms}

The algorithms below implement the RLIM realization defined in Sections~\ref{sec:rlim_family} and~\ref{sec:cover}. The minor modifications required for the non-enhanced realization of RLIM are stated after the decoder.

\subsection{Encoder}

The encoder, given in Algorithm \ref{alg1}, starts from a message index \(m\in\{0,1,\dots,2^k-1\}\). Since the selected RLIM codebook consists of the first \(2^k\) ranks of \(\mathrm{RLIM}_i(n)\), the message index \(m\) is itself the target rank. One then locates the unique Hamming-weight value \(w^\star\) whose fixed-weight subcode contains the target rank, and reconstructs the internal word bit by bit using the unranking rule of Eq. \ref{eq:general_unranking_rule}.

\begin{algorithm}[H]
\caption{Encoder}
\label{alg1}
\small
\begin{algorithmic}[1]
\Require \(i,n,k,m\), where \(m\in\{0,\dots,2^k-1\}\) is the message index
\Ensure Codeword in \(\mathrm{RLIM}_i(n,k)\)
\State \(T\gets n-i\)
\State Build or reuse the table \(F_i(\ell,s,r)\) for this value of \(T\)
\State Find the unique \(w^\star\) such that
\(\Gamma_i(T,w^\star)\le m<\Gamma_i(T,w^\star+1)\)
\State \(\rho\gets m-\Gamma_i(T,w^\star)\)
\State Initialize \(q\) as the empty word
\For{\(j=1\) to \(T\)}
    \State \(z\gets n_{\mathcal{C}_i(T,w^\star)}(q\circ 0)\)
    \If{\(\rho<z\)}
        \State \(q\gets q\circ 0\)
    \Else
        \State \(q\gets q\circ 1\); \quad \(\rho\gets \rho-z\)
    \EndIf
\EndFor
\State \Return \(0^i\circ q\)
\end{algorithmic}
\end{algorithm}

\subsection{Post-Detection Correction}

After threshold detection, each received block is represented by a binary word
\(\mathbf{y}\in\{0,1\}^{n}\).
This word need not satisfy the leading-zero and run-length constraints of the RLIM family.
For completeness, we recall the correction rule from \cite{SahinAkan2024RLIM}.
As given in Algorithm~\ref{algo2}, \(\mathbf{y}\) is mapped to a corrected word
\(\mathbf{z}\in\mathrm{RLIM}_i(n)\)
by enforcing the leading \(i\) zeros and then keeping the earliest feasible \(1\)-bits.
As shown in \cite{SahinAkan2024RLIM}, this procedure is optimal in Hamming distance for the corresponding constrained family and coincides with Viterbi decoding under the deterministic last-wins tie-break.
The corrected word \(\mathbf{z}\) is then passed to the projection-based decoder.

\begin{algorithm}[H]
\caption{Post-Detection Correction}
\label{algo2}
\small
\begin{algorithmic}[1]
\Require \(i,n,\mathbf{y}\), where \(\mathbf{y}\) is the detected binary word
\Ensure Corrected word \(\mathbf{z}\in\mathrm{RLIM}_i(n)\)
\State \(\mathbf{z}\gets 0^n\); \quad \(skip\gets i\)
\For{\(j=1\) to \(n\)}
    \If{\(skip>0\)}
        \State \(skip\gets skip-1\)
    \ElsIf{\(y_j=1\)}
        \State \(z_j\gets 1\); \quad \(skip\gets i\)
    \EndIf
\EndFor
\State \Return \(\mathbf{z}\)
\end{algorithmic}
\end{algorithm}

\subsection{Projection-Based Decoder}

\begin{table*}[!t]
\centering
\caption{Storage and computation costs of full-codebook and proposed RLIM realizations, where \(w_z\) denotes the maximum Hamming weight of a corrected word entering projection decoding, with \(w_z=O(n/i)\).}
\label{tab:complexity_summary}
\renewcommand{\arraystretch}{1.18}
\setlength{\tabcolsep}{10pt}
\footnotesize
\begin{tabular}{@{}lcc@{}}
\toprule
Operation & Full-codebook realization & Proposed realization \\
\midrule

Stored data
& \(n\cdot 2^k\) bits
& \(O(i\cdot T^2)\) integers; \(O(i\cdot T^3)\) bits \\

Preprocessing
& \(\Omega(n\cdot 2^k)\)
& \(O(i\cdot T^2)\) table updates \\

Encoding
& \(O(n)\)
& \(O(n)\) \\

Rank decoding
& \(O(k)\) codeword comparisons
& \(O(T)\) \\

Projection decoding
& \(O(w_z\cdot k)\) codeword comparisons
& \(O(w_z\cdot n)\) \\
\bottomrule
\end{tabular}
\end{table*}

After post-detection correction, the corrected word may still lie outside the selected RLIM codebook. The original RLIM workflow handles this by projection rather than by declaring a decoding failure: the decoder repeatedly erases the rightmost remaining \(1\)-bit and checks whether the resulting word belongs to the selected codebook. Membership is tested using the rank of the projected word in \(\mathrm{RLIM}_i(n)\), as shown in Algorithm~3. If the projection reaches the all-zero word, the decoder returns message index \(0\), following the original projection-to-index-\(0\) convention \cite{SahinAkan2024RLIM}.

\begin{algorithm}[H]
\caption{Projection-Based Decoder}
\small
\begin{algorithmic}[1]
\Require \(i,n,k,\mathbf{z}\), where \(\mathbf{z}\) is the corrected word
\Ensure Message index \(m\in\{0,\dots,2^k-1\}\)
\While{true}
    \State \(r\gets\operatorname{rank}_{\mathrm{RLIM}_i(n)}(\mathbf{z})\)
    \If{\(r<2^k\)}
        \State \Return \(r\)
    \EndIf
    \State Replace the rightmost \(1\) in \(\mathbf{z}\) by \(0\)
\EndWhile
\end{algorithmic}
\end{algorithm}

The non-enhanced RLIM realization in \cite{SahinAkan2024RLIM} can be implemented using the same counting, ranking, correction, and projection procedures, with only a shift in the selected rank range. Since the all-zero word is excluded, the selected codebook corresponds to ranks \(1,\dots,2^k\) rather than \(0,\dots,2^k-1\). Hence, encoding a message index \(m\in\{0,\dots,2^k-1\}\) uses the target rank \(m+1\), while decoding subtracts one from the recovered rank. Its selected codeword length must therefore be chosen such that \(|\mathcal{C}_i(n-i)|\ge 2^k+1\). If projection reaches \(0^n\), the decoder returns message index \(0\), following the projection-to-index-\(0\) convention of \cite{SahinAkan2024RLIM}.

\subsection{Complexity}

 Let \(T=n-i\), where \(n\) is the codeword length under the chosen realization. The proposed realization stores the dynamic-programming counts
\begin{equation}
F_i(\ell,s,r), \qquad 0\le \ell \le T,\quad 0\le s\le T,\quad 0\le r\le i .
\end{equation}
Thus the number of stored integer entries is
\begin{equation}
O(i\cdot T^2)=O\!\bigl(i\cdot(n-i)^2\bigr).
\label{eq:proposed_entry_complexity}
\end{equation}
Since these entries are integer counts, the number of stored bits also depends on their magnitudes. There are \(O(i\cdot T^2)\) stored count entries. Each count is at most \(2^T\), and hence can be represented using at most \(T+1\) bits. Thus, arbitrary-precision integers may be used in software, but a fixed-width implementation needs at most \(T+1\) bits per stored count. Therefore, the proposed storage is bounded by
\begin{equation}
O(i\cdot T^3)
\label{eq:proposed_bit_complexity}
\end{equation}
bits. In the numerical comparison of Fig. 2 (a), the proposed storage is computed more tightly by summing the actual bit lengths of all stored integers.

The previous full-codebook realization stores all \(2^k\) selected length-\(n\) codewords and requires $n\cdot 2^k $ bits. The proposed realization stores only the counting tables used for ranking and unranking. For fixed order \(i\) and the shortest admissible selected length, \(k\le T\le(i+1)k\), and thus \(T=\Theta(k)\). Hence full codebook storage grows exponentially in \(k\), whereas the proposed table storage grows polynomially in \(k\). Table~\ref{tab:complexity_summary} summarizes these costs. The proposed method reduces storage, preprocessing, rank decoding, and projection decoding. Encoding changes from direct lookup to unranking; this is the cost of avoiding stored codewords.

\begin{figure*}[!t]
\centering

\newlength{\COMPHOVERLAP}
\setlength{\COMPHOVERLAP}{3mm}

\newlength{\COMPPANELW}
\setlength{\COMPPANELW}{\dimexpr 0.25\textwidth + 0.75\COMPHOVERLAP \relax}

\begin{tabular}{@{}c@{\hspace{-\COMPHOVERLAP}}c@{\hspace{-\COMPHOVERLAP}}c@{\hspace{-\COMPHOVERLAP}}c@{}}
\multicolumn{4}{@{}c@{}}{
  \includegraphics[width=\dimexpr 4\COMPPANELW-3\COMPHOVERLAP\relax]{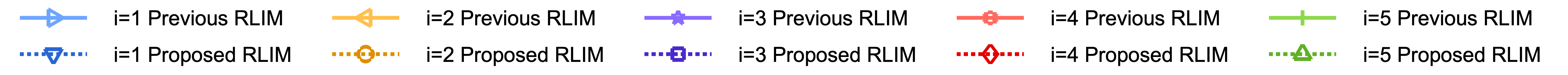}
}\\[-1mm]

\subcaptionbox{Storage requirement}{\includegraphics[width=\COMPPANELW]{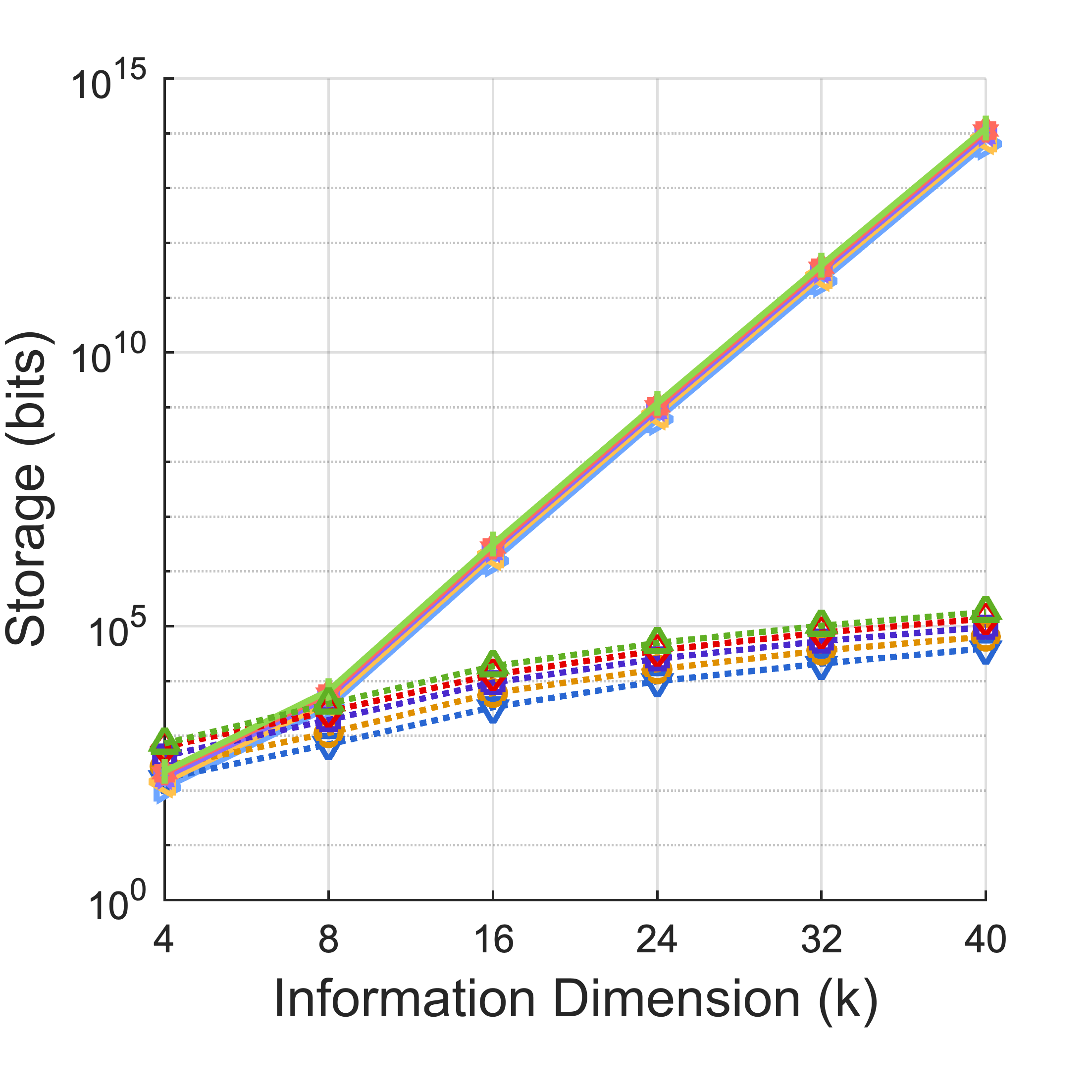}} &
\subcaptionbox{Preprocessing time}{\includegraphics[width=\COMPPANELW]{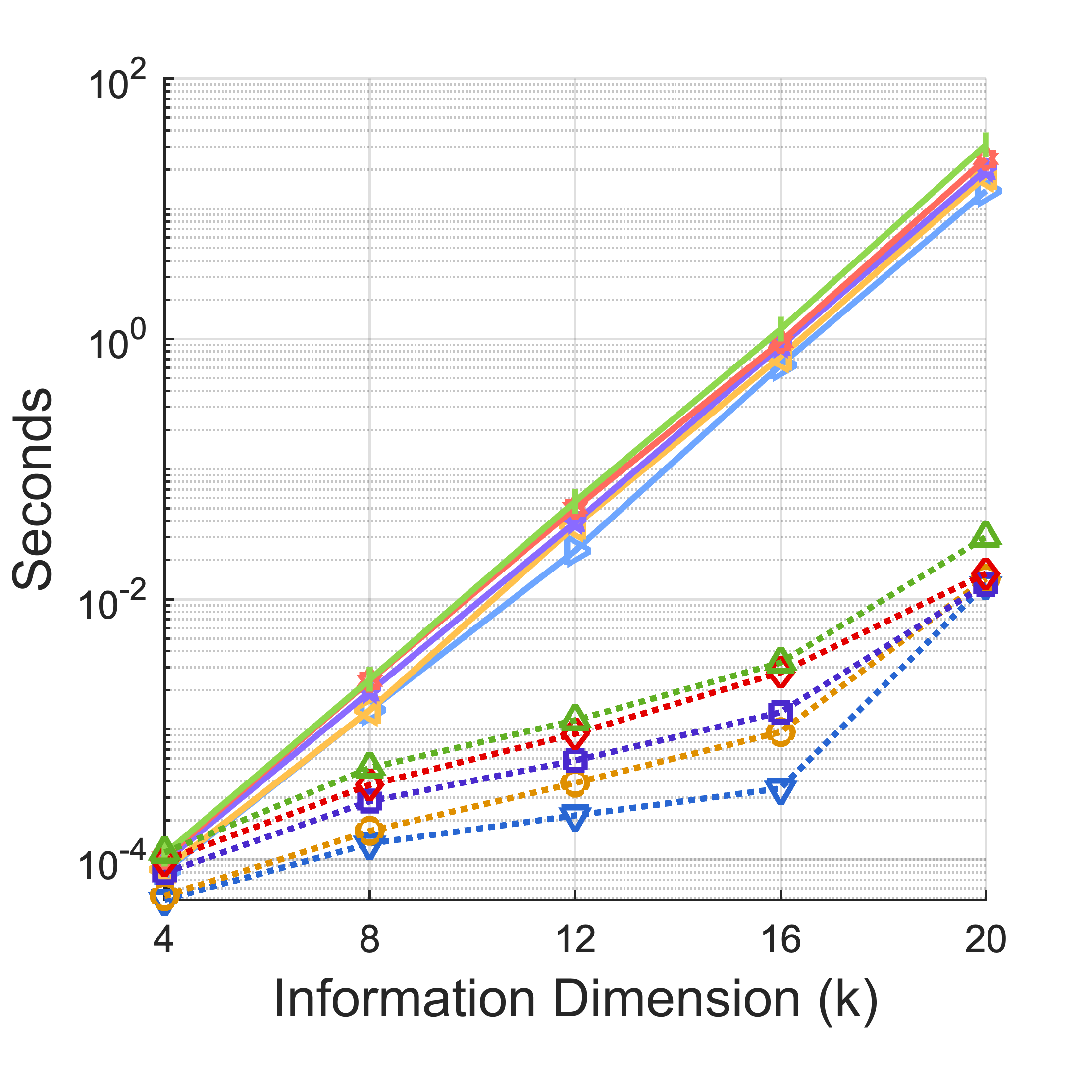}} &
\subcaptionbox{Encoding time}{\includegraphics[width=\COMPPANELW]{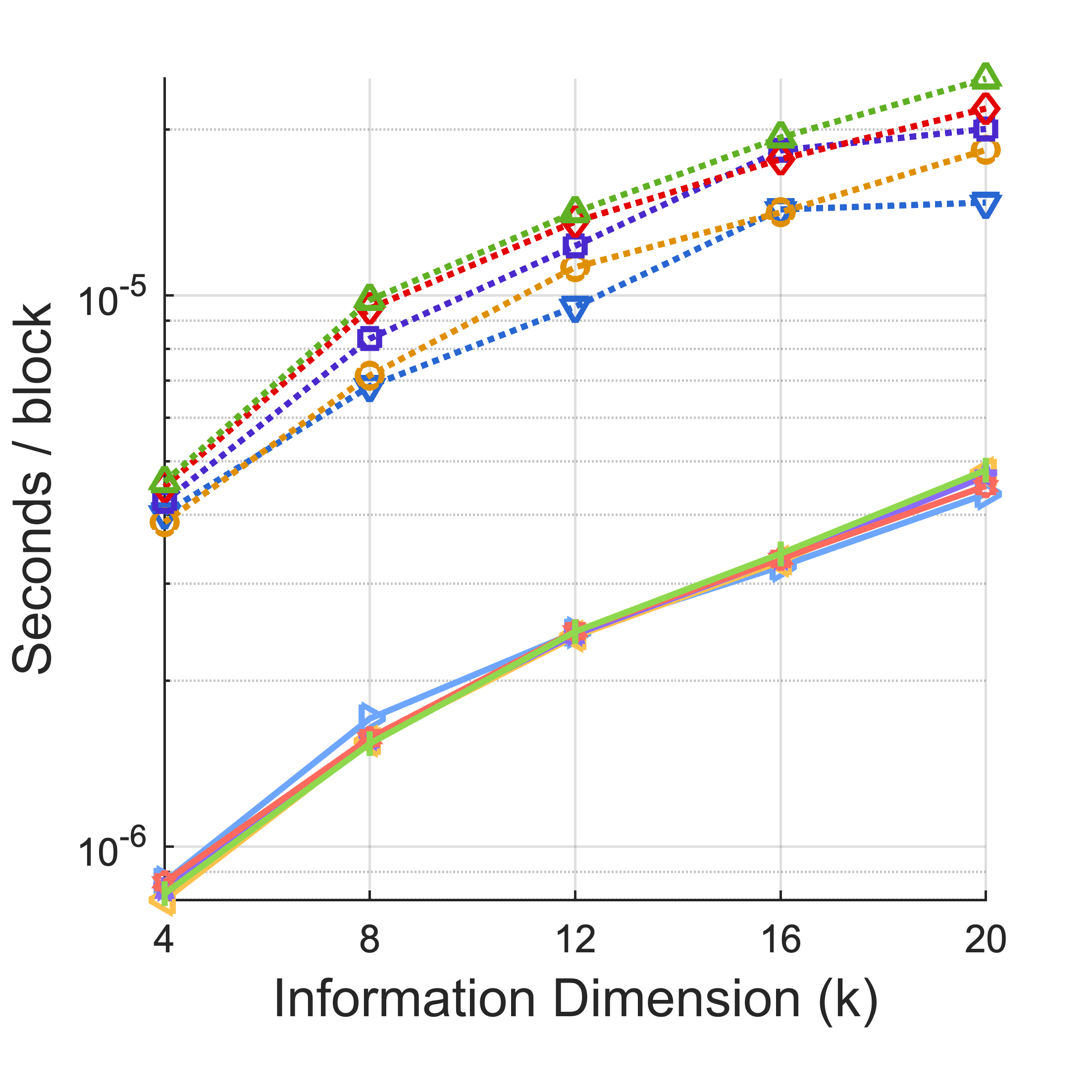}} &
\subcaptionbox{Rank decoding time}{\includegraphics[width=\COMPPANELW]{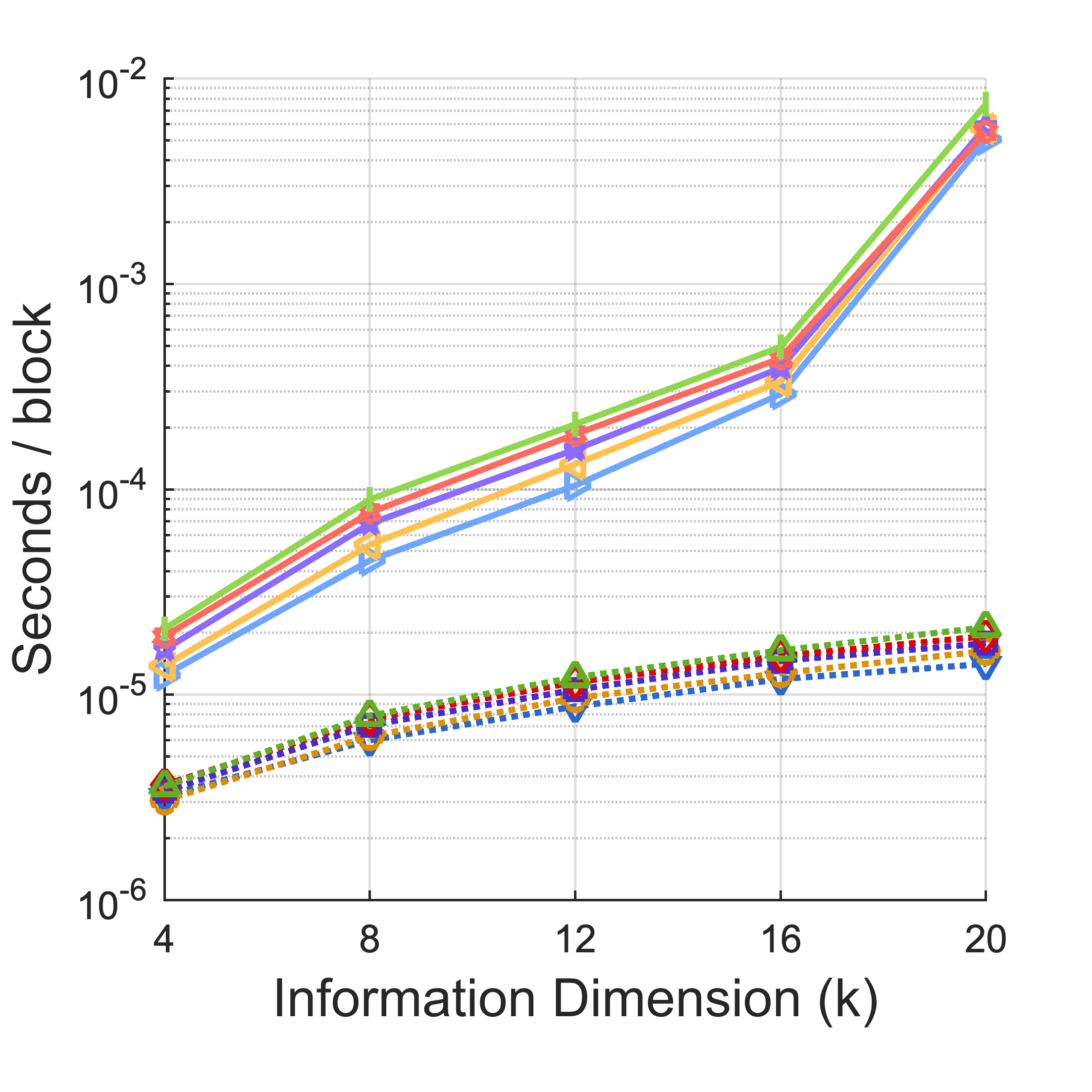}}
\end{tabular}

\caption{Storage and runtime comparison between the previous full-codebook RLIM realization and the proposed low-storage realization. Encoding and rank-decoding times are average per-block times over random message blocks; preprocessing time is measured once for each \((i,k)\). Proposed storage is the sum of the bit lengths of the stored integer counts. For each \((i,k)\), both realizations use the same selected RLIM codebook and the same shortest admissible codeword length \(n\); only the realization differs.}
\label{fig:complexity_storage_runtime}
\end{figure*}

Fig.~\ref{fig:complexity_storage_runtime} compares the two realizations for \(i=1,\dots,5\). Panel (a) shows that full codebook storage grows exponentially in \(k\), while the proposed counting tables remain small over the tested range. For example, at \(i=3\) and \(k=40\), full codebook storage is about \(12.1\) TB, whereas the proposed tables require about \(12.1\) kB. Panel (b) shows the corresponding preprocessing reduction. Panel (c) shows the expected encoding trade-off between lookup and unranking. Panel (d) shows the corresponding reduction in decoding time, obtained by replacing stored-codebook search with direct rank computation.

This reduction removes a major implementation bottleneck for RLIM in resource-constrained MC settings. RLIM was already attractive at the code-design level, but full codebook storage made larger selected codebooks difficult to use and unsuitable for resource-constrained transmitters and receivers. The proposed realization removes this lookup-table barrier, replacing exponential codebook storage with polynomial-size indexing tables for ranking and unranking. The present work establishes these gains at the algorithmic level through storage and runtime analysis. A physical implementation on a nanoscale MC node is beyond the scope of this paper and will additionally depend on the available digital memory, arithmetic precision, processing latency, and energy budget of the target platform. Future work will investigate these hardware-level requirements on resource-constrained MC transmitter and receiver platforms.

For completeness, the competing codes used in Section~VI also have low implementation complexity. In the configurations considered here, the fixed \((15,11)\) Hamming and ISI-free \((4,2,1)\) codes perform constant work per short block, the MEC decoder searches only \(16\) codewords, and the Reed--Solomon baseline uses a fixed two-parity-byte algebraic encoder and decoder. Their processing cost therefore grows linearly with the processed bit stream. The original RLIM realization likewise had linear scaling for encoding and rank decoding, but required storage of the full size-\(2^k\) codebook and decoding through stored-codeword comparisons. The proposed realization preserves this linear scaling while removing the exponential codebook storage and replacing stored-codeword search with counting-table operations. For \(i=3\) and \(k=16\), the implemented counting tables require only \(9{,}359\) bits, compared with \(2{,}424{,}832\) bits for the full RLIM codebook.

\captionsetup[subfigure]{labelformat=parens}
\captionsetup[subfigure]{justification=centering, singlelinecheck=false}

\captionsetup[subfigure]{labelformat=parens}
\captionsetup[subfigure]{justification=centering, singlelinecheck=false}

\begin{figure*}[t]
\centering

\newlength{\HOVERLAP}
\setlength{\HOVERLAP}{3mm}

\newlength{\PANELW}
\setlength{\PANELW}{\dimexpr 0.25\textwidth + 0.75\HOVERLAP \relax}

\newlength{\ROWGAP}
\setlength{\ROWGAP}{3mm}

\makebox[\textwidth][c]{%
\begin{tabular}{@{}c@{\hspace{-\HOVERLAP}}c@{\hspace{-\HOVERLAP}}c@{\hspace{-\HOVERLAP}}c@{}}
\multicolumn{4}{@{}c@{}}{
  \includegraphics[width=\dimexpr 3.8\PANELW-3\HOVERLAP\relax]{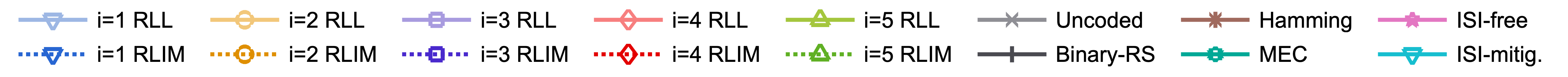}
}\\[-1mm]

  \subcaptionbox{\(t_{s,0}=0.2\) \(\mathrm{s}\)}{\includegraphics[width=\PANELW]{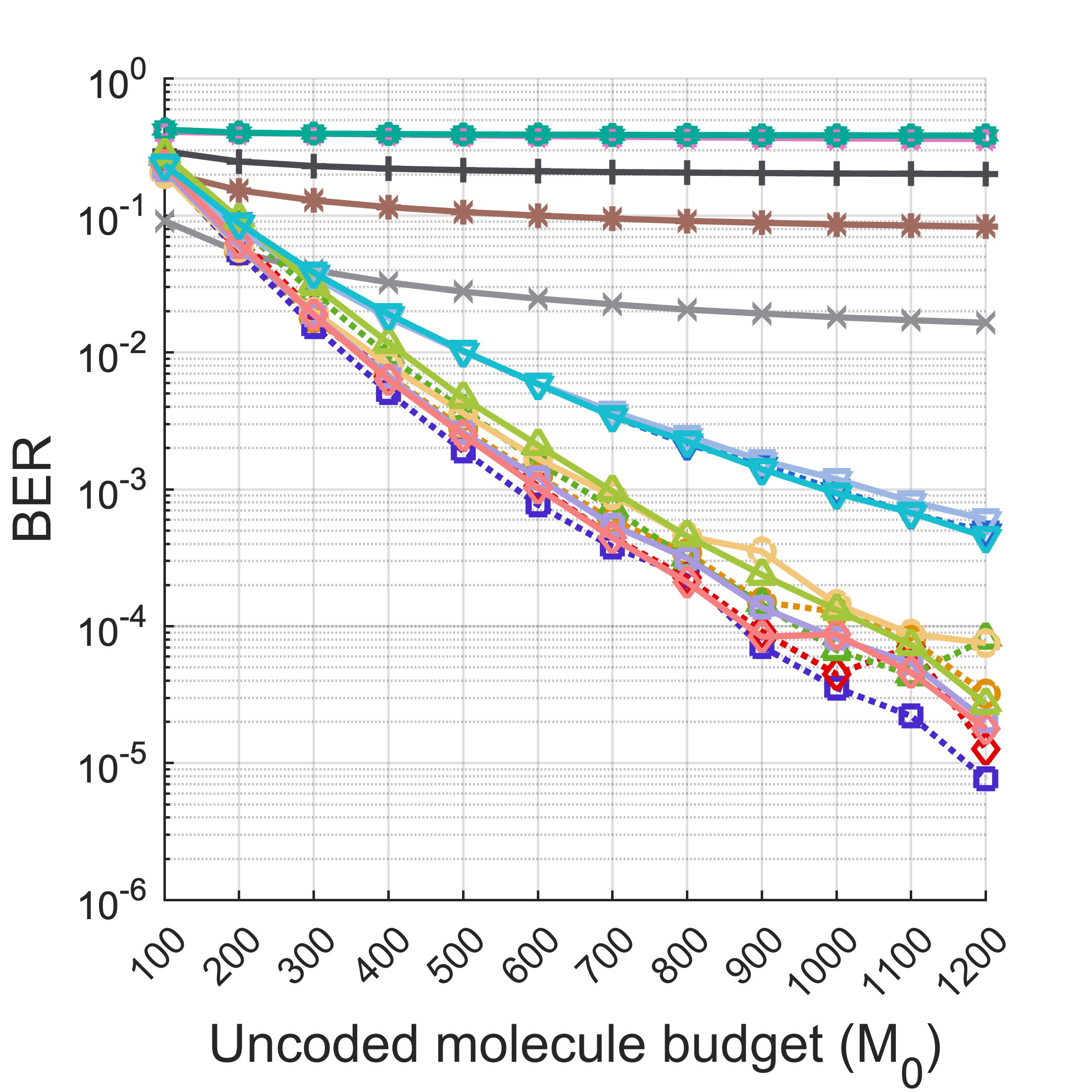}} &
  \subcaptionbox{\(M_0=100\)}{\includegraphics[width=\PANELW]{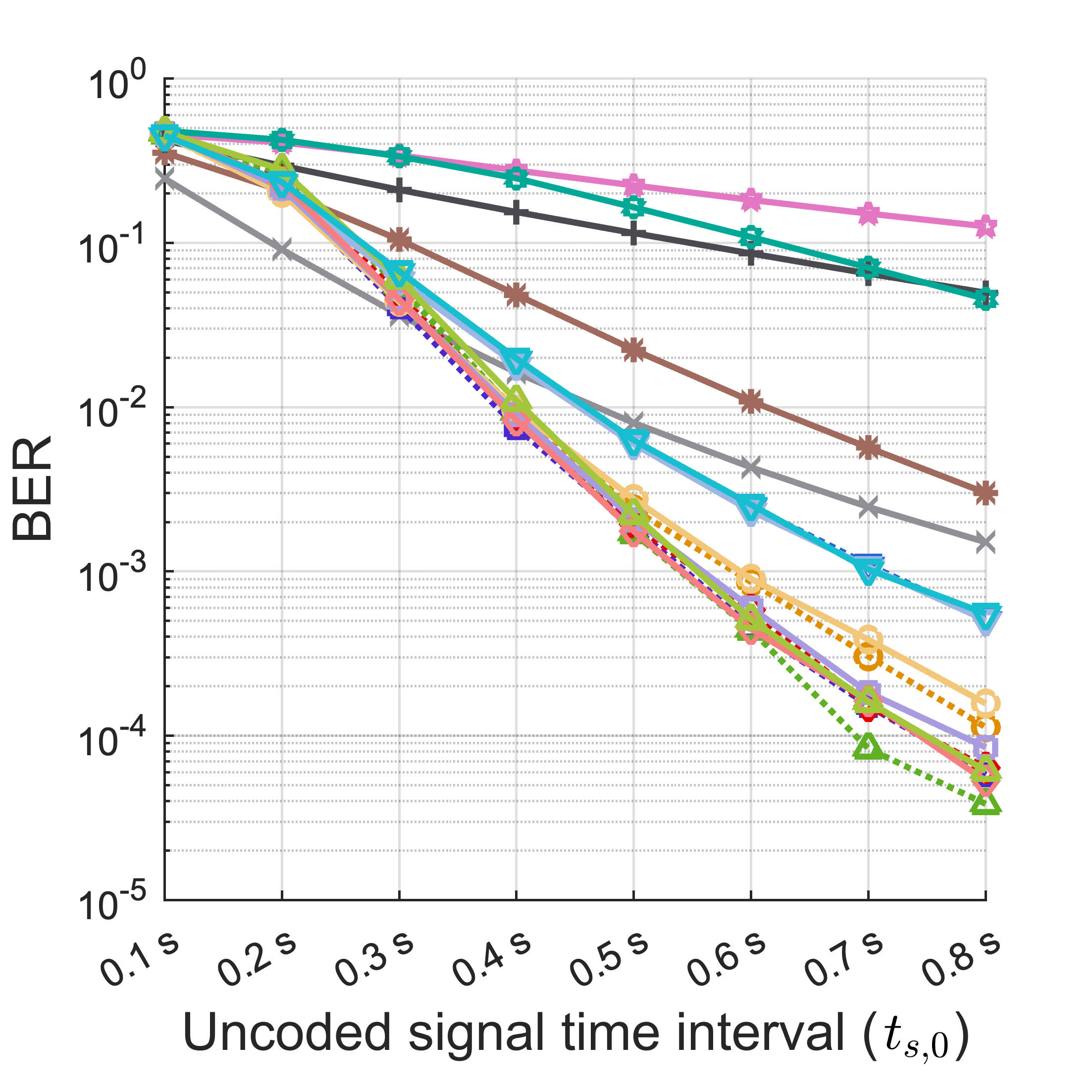}} &
  \subcaptionbox{\(M_0=500\), \(t_{s,0}=0.2\) \(\mathrm{s}\)}{\includegraphics[width=\PANELW]{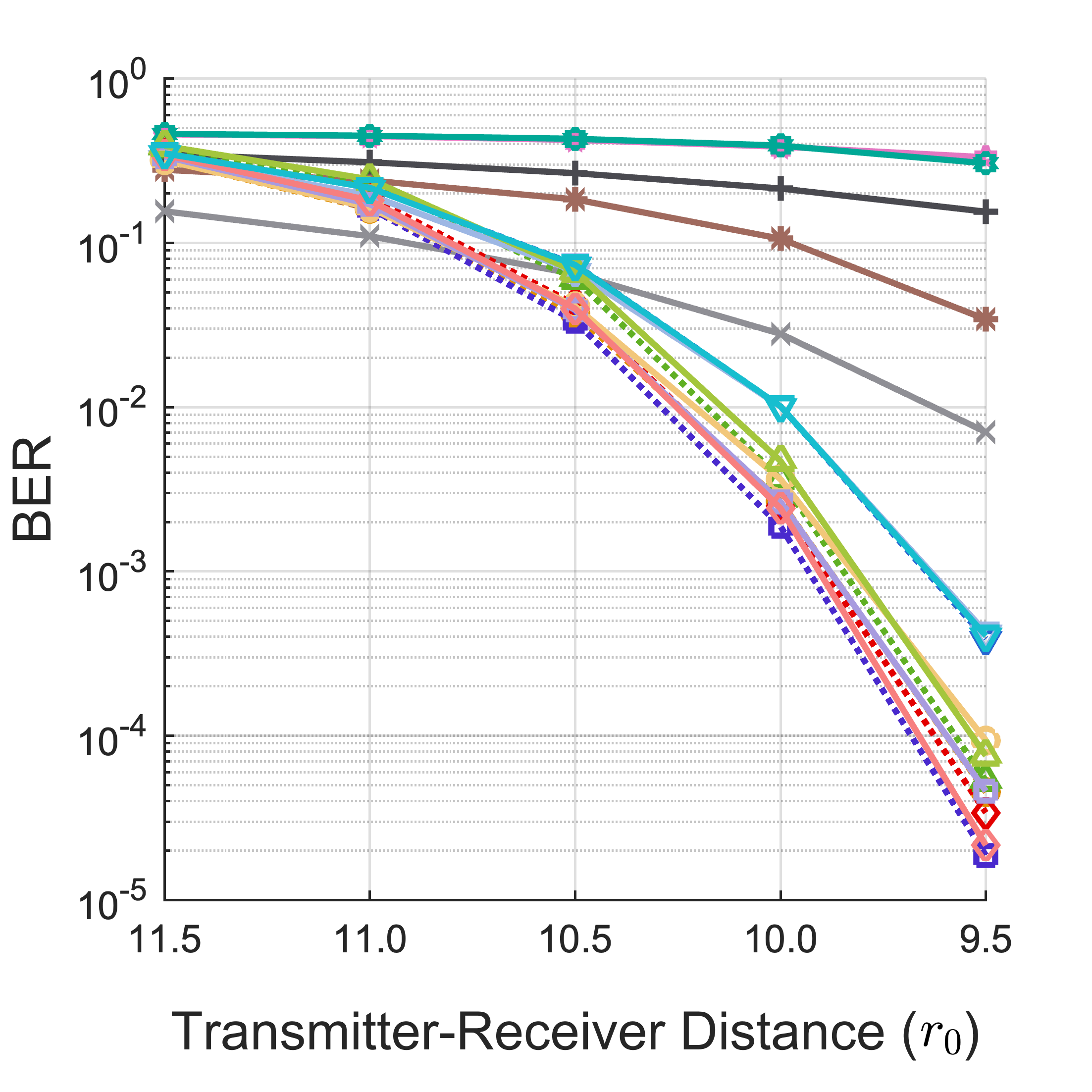}} &
  \subcaptionbox{\(M_0=500\), \(t_{s,0}=0.2\) \(\mathrm{s}\)}{\includegraphics[width=\PANELW]{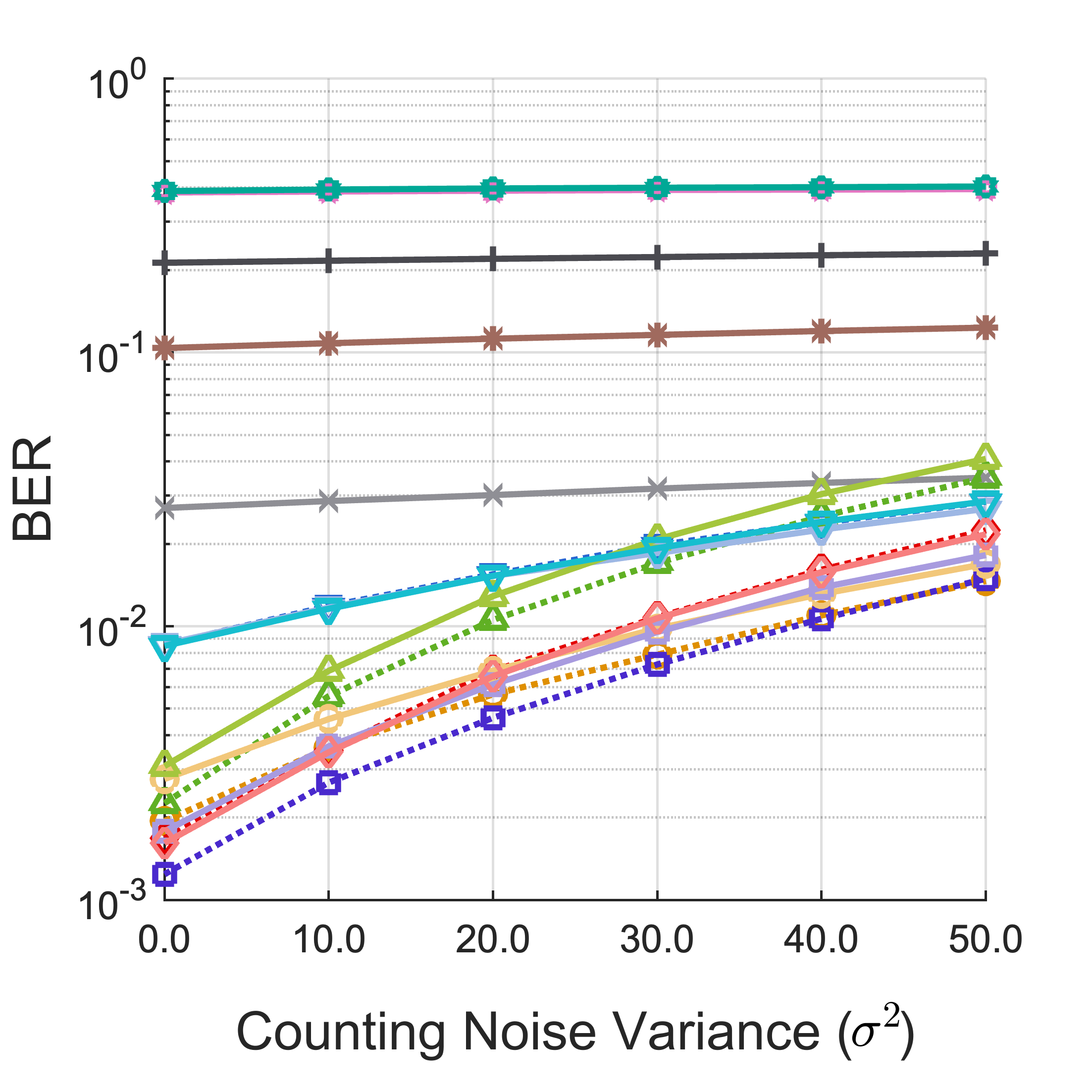}}

  \\[\ROWGAP]

  \subcaptionbox{\(M_0=500\), \(t_{s,0}=0.2\) \(\mathrm{s}\)}{\includegraphics[width=\PANELW]{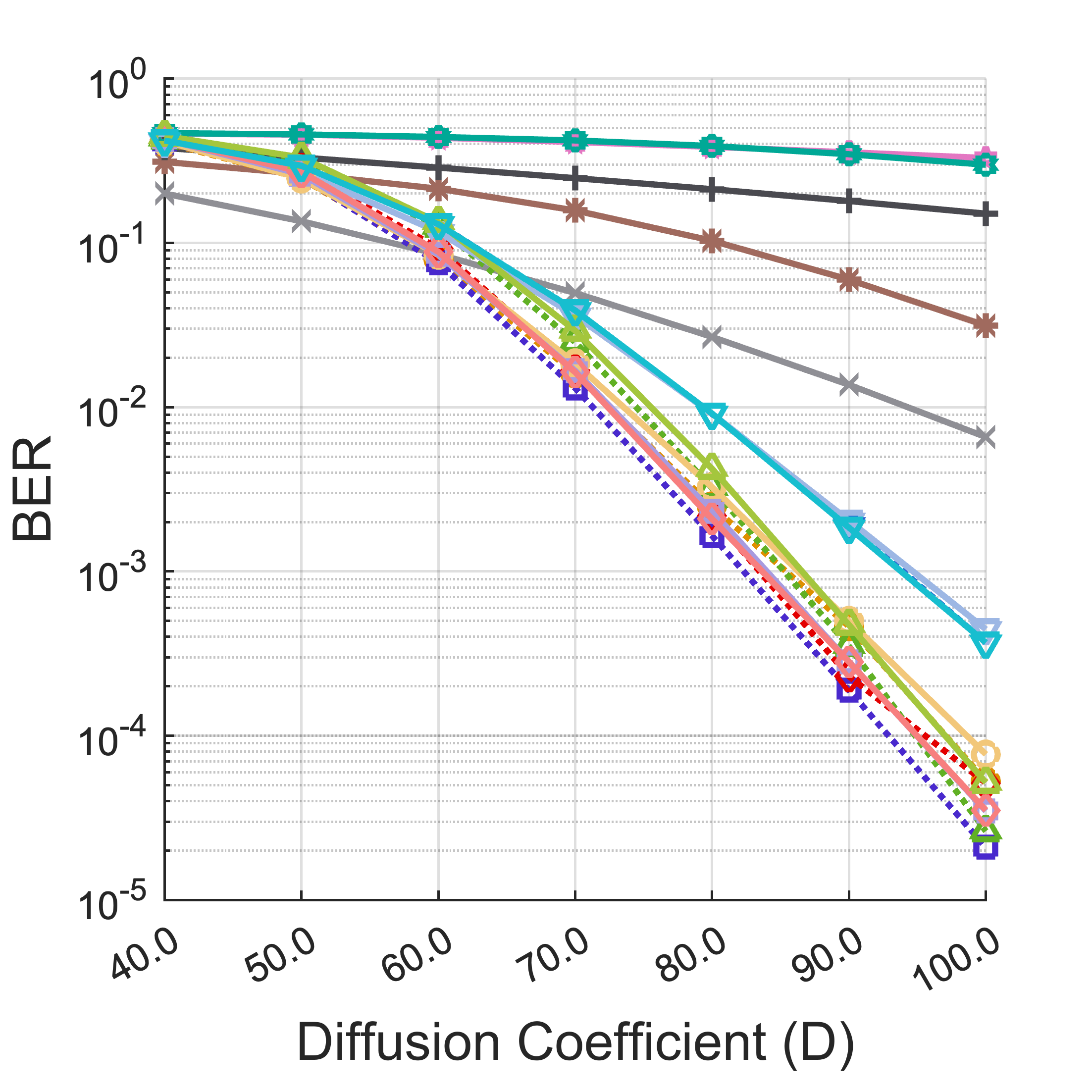}} &
  \subcaptionbox{\(M_0=800\), \(t_{s,0}=0.2\) \(\mathrm{s}\)}{\includegraphics[width=\PANELW]{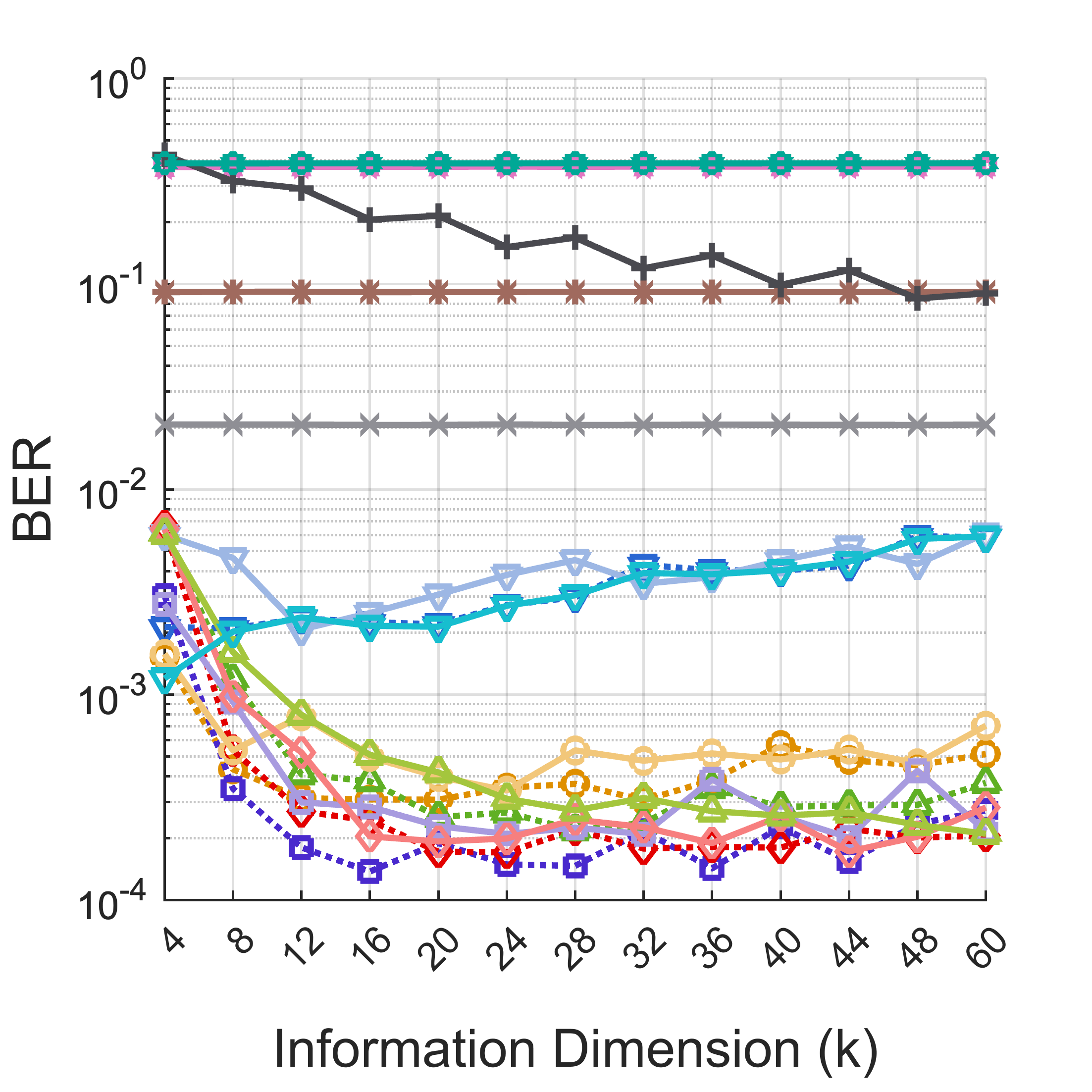}} &
  \subcaptionbox{\(M_0=500\), \(t_{s,0}=0.25\) \(\mathrm{s}\)}{\includegraphics[width=\PANELW]{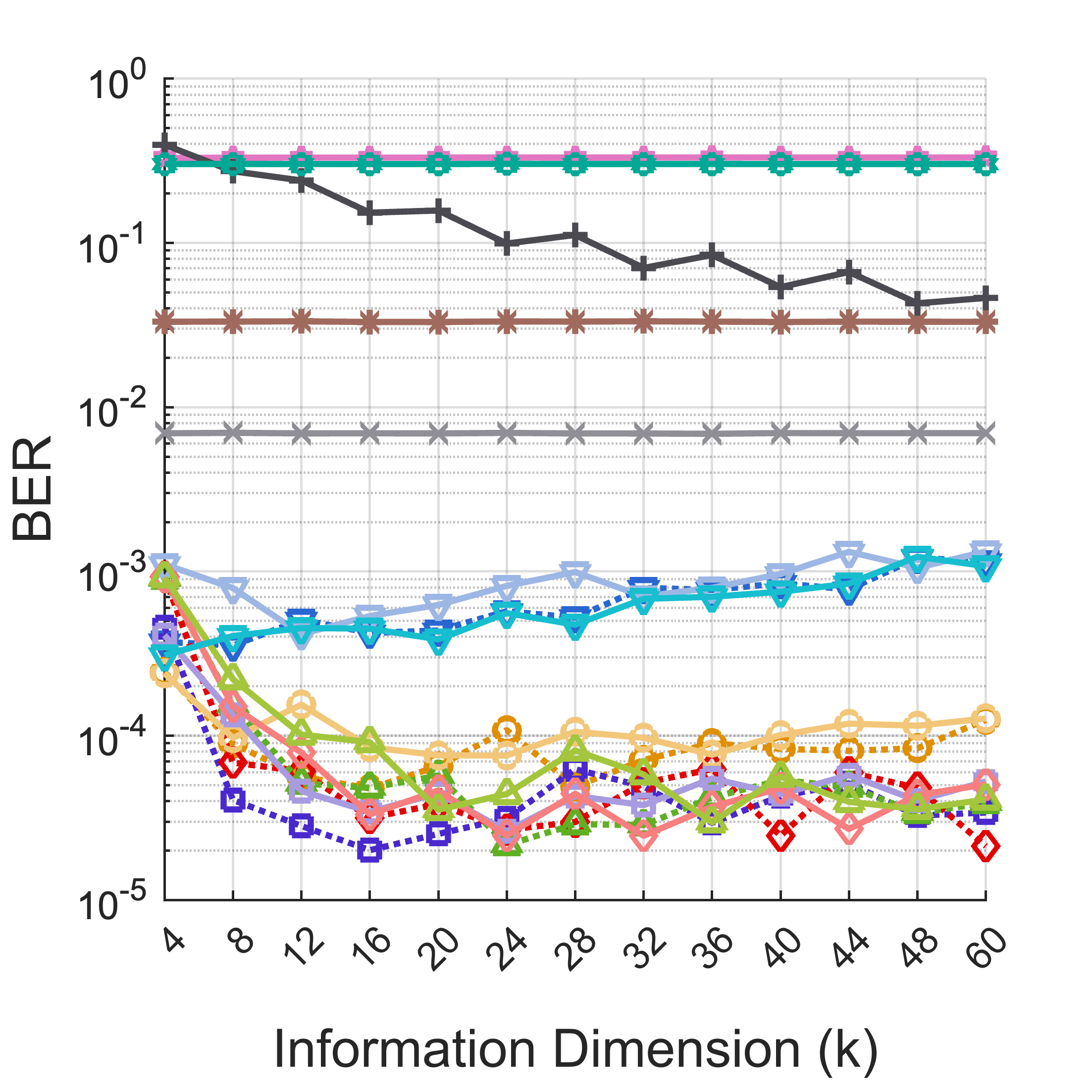}} &
  \subcaptionbox{\(M_0=100\), \(t_{s,0}=0.8\) \(\mathrm{s}\)}{\includegraphics[width=\PANELW]{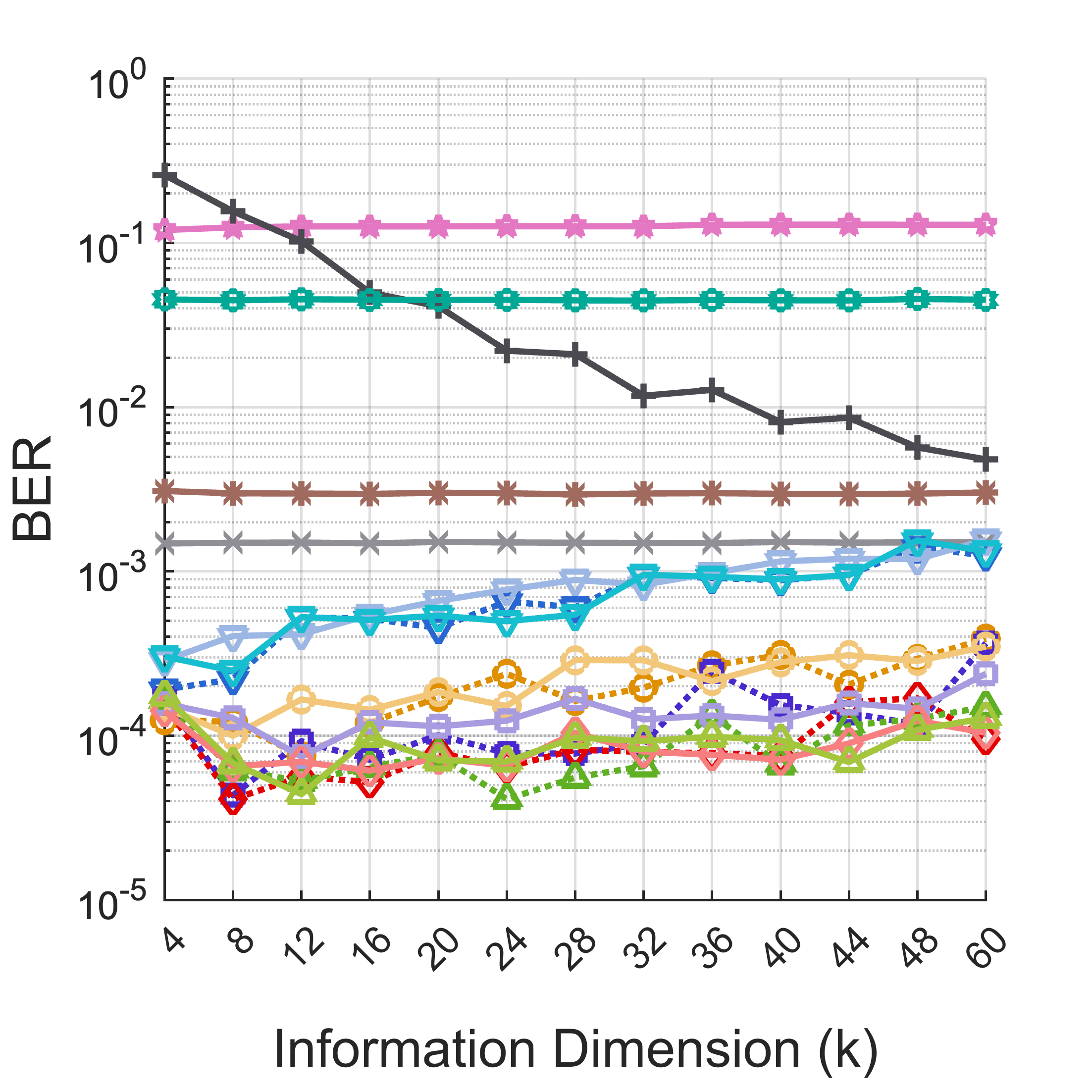}} \\
\end{tabular}
}

\caption{Bit-error-rate values across methods. Unless stated otherwise, \(D=79.4~\mu\mathrm{m}^2/\mathrm{s}\), \(r_R=5~\mu\mathrm{m}\), \(r_0=10~\mu\mathrm{m}\), \(k=16\), \(\sigma^2=5\), and \(I_0=100\). For scheme \(m\), the channel memory is \(I_m=\lceil I_0n_m/K_m\rceil\), preserving the retained physical channel-memory duration after signal-interval normalization. Each plotted BER is averaged over five independent repetitions.}
\label{fig:all_16_shared_legend_3}
\end{figure*}

\section{Simulation Framework and Numerical Results}
\label{sec:applications}

\subsection{Simulation Framework}

The compared methods are uncoded transmission, the \((15,11)\) Hamming code \cite{Hamming1950}, the ISI-free \((4,2,1)\) code \cite{Shih2013}, a Reed--Solomon baseline with two parity bytes \cite{ReedSolomon1960,Dissanayake2017RS}, a minimum-energy code (MEC) \cite{BaiLeeson2014}, the ISI-mitigating code \cite{Kislal2020}, RLIM codebooks of orders \(i=1,2,3,4,5\), and the corresponding lexicographic \((i,\infty)\)-RLL codebooks. The MEC baseline uses \(M_{\mathrm{MEC}}=16\) codewords, i.e., \(4\) information bits per MEC block, and minimum Hamming distance \(d_{\mathrm{MEC}}=3\).

The normalization follows the whole-codebook \(1\)-bit counts. Let \(t_{s,0}\) and \(M_0\) denote the signal interval and molecule count used by uncoded transmission for one information bit. For scheme \(m\), let \(K_m\) be the number of information bits per block, \(n_m\) the coded block length, and \(W_m\) the total number of \(1\)-bits across its full codebook. Let \(W_{0,m}\) denote the total number of \(1\)-bits in the corresponding uncoded codebook with \(K_m\) information bits.  Following \cite{ldp, normalization}, the signal interval and molecule count of scheme \(m\) are chosen as
\begin{equation}
t_{s,m}=t_{s,0}\cdot\frac{K_m}{n_m},
\label{eq:norm_ts}
\end{equation}
and
\begin{equation}
M_m=
\left\lfloor
M_0\cdot\frac{W_{0,m}}{W_m}
\right\rceil .
\label{eq:norm_M}
\end{equation}Thus one encoded block of scheme \(m\) occupies the same total time as \(K_m\) uncoded information bits, and its full codebook uses approximately the same total molecule budget as the corresponding uncoded \(K_m\)-bit codebook.

Because \(t_{s,m}\) varies with the coding rate, using the same number of channel taps for all schemes would result in different retained memory durations in physical time. We therefore set
\begin{equation}
I_m=\left\lceil I_0 \cdot \frac{n_m}{K_m}\right\rceil .
\label{eq:norm_I}
\end{equation}
The ceiling ensures \(I_m \cdot  t_{s,m}\ge I_0 t_{s,0}\), so that no coded scheme is simulated with a shorter physical channel history than the uncoded reference. 

For each operating point, independent training and test sets are transmitted consecutively, so that inter-block ISI is retained. The training sequence has \(110{,}880\) information bits, and the test sequence has \(2{,}217{,}600\) information bits. Both sequences are generated independently at random. The complete training-and-test procedure is repeated independently five times for every method and operating point. The simulation parameters are provided in the caption of Fig.~\ref{fig:all_16_shared_legend_3}.\footnote{The default channel settings \(D=79.4~\mu\mathrm{m}^2/\mathrm{s}\), \(r_R=5~\mu\mathrm{m}\), and \(r_0=10~\mu\mathrm{m}\) follow the commonly used insulin-molecule benchmark in diffusion-based MC, where \(D\) is the diffusion coefficient of human insulin molecules, \(r_R\) is the receiver radius, and \(r_0\) is the transmitter--receiver distance \cite{ISI_mitigating_methods_2015}.} For each method \(m\), the received count sequence is simulated using \eqref{eq:mc_received}.

The choices \(\sigma^2=5\) and \(I_0=100\) can also be related directly to the channel response. For \(D=79.4~\mu\mathrm{m}^2/\mathrm{s}\), \(r_R=5~\mu\mathrm{m}\), \(r_0=10~\mu\mathrm{m}\), and \(t_{s,0}=0.2~\mathrm{s}\), Eq.~\eqref{eq:mc_F} gives \(p_1=F(0.2)=0.18748\). Hence, at \(M_0=500\), an isolated transmitted \(1\)-symbol contributes on average \(M_0\cdot p_1=93.74\) molecules in the first interval. The corresponding first-interval molecule count has variance \(M_0\cdot p_1\cdot(1-p_1)=76.17\), whereas the added receiver noise has variance \(\sigma^2=5\). Thus, \(\sigma^2=5\) increases the first-interval count variance by only about \(6.6\%\), corresponding to an increase in standard deviation from \(8.73\) to \(9.01\) molecules. It therefore represents a moderate receiver-noise level that perturbs, rather than dominates, the stochastic variation due to molecular absorption. Fig.~\ref{fig:all_16_shared_legend_3}(d) further sweeps \(\sigma^2\) from \(0\) to \(50\), so the reported conclusions are not exclusively tied to \(\sigma^2=5\).

For the channel memory, \(I_0=100\) at \(t_{s,0}=0.2~\mathrm{s}\) corresponds to a \(20~\mathrm{s}\) retained channel-memory window. Under the same channel geometry, \(F(20)=0.46465\), while \(F(\infty)=r_R/r_0=0.5\). Hence, the retained window contains \(F(20)/F(\infty)=92.93\%\) of the molecules that will eventually be absorbed. The probability associated with the 100th tap is \(p_{100}=F(20)-F(19.8)=1.776\times10^{-4}\), so a single emission with \(M_0=500\) contributes on average only \(500\cdot p_{100}=0.0888\) molecules through this tap. Doubling the reference memory to \(I_0=200\) would extend the retained window to \(40~\mathrm{s}\) and increase the captured fraction of eventually absorbed molecules only to \(94.998\%\), while doubling the number of retained taps. Together with the normalization in Eqs.~\eqref{eq:norm_ts} and \eqref{eq:norm_I}, $I_0=100$ provides a sufficiently long physical channel-memory window across the compared schemes.

For each scheme and operating point, the static threshold is selected using a training transmission. The superscript ``trn'' denotes training quantities. Let \(N_{m,t}^{\mathrm{trn}}\) be the received molecule count in the \(t\)-th transmitted symbol interval of scheme \(m\) during training. For a candidate threshold \(\tau\), the hard threshold output is
\begin{equation}
\widehat{y}_{m,t}^{\mathrm{trn}}(\tau)=
\begin{cases}
1, & N_{m,t}^{\mathrm{trn}}\ge \tau,\\
0, & N_{m,t}^{\mathrm{trn}}< \tau.
\end{cases}
\label{eq:threshold_rule}
\end{equation}
 The vector \(\widehat{\mathbf y}_{m}^{\mathrm{trn}}(\tau)\) is the threshold-detected transmitted-bit sequence before any code-specific post-processing.

The candidate thresholds are the integers
\begin{equation}
\tau\in\{0,1,\dots,\max\{0,\max_t N_{m,t}^{\mathrm{trn}}\}\}.
\label{eq:threshold_candidates}
\end{equation}
For each candidate threshold, the detected sequence \(\widehat{\mathbf y}_{m}^{\mathrm{trn}}(\tau)\) is passed through the complete receiver of scheme \(m\), including any scheme-specific decoding step and, where applicable, correction and projection. Let
\begin{equation}
\widehat{\mathbf b}_{m}^{\mathrm{trn}}(\tau)
=
\bigl(\widehat b_{m,1}^{\mathrm{trn}}(\tau),\dots,
\widehat b_{m,|\mathbf b^{\mathrm{trn}}|}^{\mathrm{trn}}(\tau)\bigr)
\end{equation}
denote the resulting decoded information-bit sequence, and let
\(\mathbf b^{\mathrm{trn}}\) denote the transmitted training information-bit sequence. The selected threshold is
\begin{equation}
\tau_m^\star
\in
\arg\min_{\tau}
\sum_{u=1}^{|\mathbf{b}^{\mathrm{trn}}|}
\mathbf{1}
\!\left[
b_u^{\mathrm{trn}}
\neq
\widehat{b}^{\mathrm{trn}}_{m,u}(\tau)
\right].
\label{eq:threshold_argmin}
\end{equation}
The selected threshold \(\tau_m^\star\) is then fixed and applied to an independent test transmission. Let \(\mathbf b^{\mathrm{te}}\) denote the transmitted test information-bit sequence, and let \(\widehat{\mathbf b}^{\mathrm{te}}_m(\tau_m^\star)\) denote the corresponding decoded test information-bit sequence for scheme \(m\). The BER for a single repetition is
\begin{equation}
\mathrm{BER}_m
=
\frac{1}{|\mathbf b^{\mathrm{te}}|}
\sum_{u=1}^{|\mathbf b^{\mathrm{te}}|}
\mathbf{1}\!\left[
b_u^{\mathrm{te}}
\neq
\widehat b^{\mathrm{te}}_{m,u}(\tau_m^\star)
\right].
\label{eq:test_ber}
\end{equation}
The reported BER is the arithmetic mean of the BER values obtained over the five independent repetitions. If several thresholds minimize the training error, ties are resolved using the convention of \cite{SahinAkan2024RLIM}.\footnote{Let \(\mathcal{T}_m^\star\) be the set of minimizing thresholds and set \(\tau_{\mathrm{mid}}=\lfloor(\min\mathcal{T}_m^\star+\max\mathcal{T}_m^\star)/2\rfloor\). The selected threshold is the smallest \(\tau\in\mathcal{T}_m^\star\) satisfying \(|\tau-\tau_{\mathrm{mid}}|\le 1\); if no such threshold exists, \(\min\mathcal{T}_m^\star\) is used.}

\subsection{Numerical Results}
\label{sec:results}

Fig.~\ref{fig:all_16_shared_legend_3} reports eight BER sweeps. Panels (a)--(e) vary, respectively, the uncoded molecule budget \(M_0\), the uncoded signal interval \(t_{s,0}\), the transmitter--receiver distance \(r_0\), the receiver counting-noise variance \(\sigma^2\), and the diffusion coefficient \(D\), with \(k=16\). Panels (f)--(h) vary the information dimension \(k\) in three representative regimes. The conclusions below are based on persistent trends across operating points and on the aggregate comparisons in Tables~\ref{tab:rlim-vs-rll-elementwise} and \ref{tab:overall_winner}.

In the fixed-\(k\) sweeps of Fig.~\ref{fig:all_16_shared_legend_3}(a)--(e), the best RLIM and RLL variants generally move away from the classical baselines as the channel becomes less severe. Increasing \(M_0\), increasing \(t_{s,0}\), decreasing \(r_0\), or increasing \(D\) improves the BER of the strongest constrained-code variants by several orders of magnitude. The counting-noise sweep in Fig.~\ref{fig:all_16_shared_legend_3}(d) is more difficult for all methods, but the best RLIM curves remain below the classical coding baselines over the tested range.

The \(k\)-sweeps in Fig.~\ref{fig:all_16_shared_legend_3}(f)--(h) illustrate the practical consequence of the proposed realization. Larger selected RLIM codebooks can now be constructed, encoded, and decoded without storing the full codebook. The original full-codebook RLIM implementation was evaluated at \(k=16\), whereas the proposed realization enables substantially larger information dimensions to be explored directly. The newly accessible \(k>16\) points improve the \(k=16\) BER of several individual RLIM orders. In panel (g), RLIM$_4$ improves from \(3.166\times10^{-5}\) at \(k=16\) to \(2.128\times10^{-5}\) at \(k=60\), while RLIM$_5$ improves from \(4.816\times10^{-5}\) at \(k=16\) to \(2.165\times10^{-5}\) at \(k=24\). In panel (f), RLIM$_4$ improves from \(2.441\times10^{-4}\) at \(k=16\) to \(1.706\times10^{-4}\) at \(k=24\), while RLIM$_5$ improves from \(3.787\times10^{-4}\) at \(k=16\) to \(2.186\times10^{-4}\) at \(k=28\). In panel (h), RLIM$_5$ improves from \(6.457\times10^{-5}\) at \(k=16\) to \(4.113\times10^{-5}\) at \(k=24\).

Thus, although increasing \(k\) does not universally reduce BER, removing the full-codebook storage bottleneck exposes previously impractical larger-dimensional operating points that can improve the performance of individual RLIM orders relative to the \(k=16\) regime. The \(k\)-sweeps also reveal a non-monotonic tradeoff: beyond a favorable information dimension for a given RLIM order and channel condition, further increasing \(k\) can degrade BER, with one contributing factor being that a block-decoding error can affect multiple information bits within the \(k\)-bit message block.

\begin{table}[!t]
\centering
\caption{Elementwise BER comparison of RLIM$_i$ and RLL$_i$ at the same order.}
\label{tab:rlim-vs-rll-elementwise}
\renewcommand{\arraystretch}{1.05}
\setlength{\tabcolsep}{5pt}
\footnotesize
\begin{tabular}{c|cc|cc|c}
\hline
\(i\) & \multicolumn{2}{c|}{RLIM$_i$ wins} & \multicolumn{2}{c|}{RLL$_i$ wins} & ties \\
\cline{2-3}\cline{4-5}
 & count & mean \((\mathrm{RLL}/\mathrm{RLIM})\) & count & mean \((\mathrm{RLIM}/\mathrm{RLL})\) & \\
\hline
1 & 41 & \(1.385\times\) & 36 & \(1.098\times\) & 0 \\
2 & 62 & \(1.374\times\) & 15 & \(1.146\times\) & 0 \\
3 & 65 & \(1.527\times\) & 12 & \(1.284\times\) & 0 \\
4 & 31 & \(1.353\times\) & 46 & \(1.182\times\) & 0 \\
5 & 61 & \(1.379\times\) & 16 & \(1.400\times\) & 0 \\
\hline
\end{tabular}
\end{table}

\begin{table}[t]
\centering
\caption{Methods that attain at least one best result across the 77 simulated operating points. A unique win means that the method attains the smallest BER at that operating point without a tie.}
\label{tab:overall_winner}
\renewcommand{\arraystretch}{1.05}
\setlength{\tabcolsep}{4pt}
\footnotesize
\begin{tabular}{lccc}
\hline
Method & Unique wins & Tied-best & Mean dominance \\
\hline
RLIM$_3$ & 37 & 0 & \(1.282\times\) \\
RLIM$_4$ & 10 & 0 & \(1.224\times\) \\
RLIM$_5$ & 10 & 0 & \(1.268\times\) \\
Uncoded & 8 & 0 & \(1.703\times\) \\
RLL$_4$ & 4 & 0 & \(1.181\times\) \\
RLL$_5$ & 4 & 0 & \(1.153\times\) \\
RLIM$_2$ & 2 & 0 & \(1.078\times\) \\
RLL$_2$ & 1 & 0 & \(1.018\times\) \\
ISI-mitigating & 1 & 0 & \(1.262\times\) \\
\hline
\end{tabular}
\end{table}

Table~\ref{tab:rlim-vs-rll-elementwise} compares RLIM and lexicographic RLL point by point at the same order. The minimum-weight selection used by RLIM is favorable for most tested orders. For \(i=2\), \(i=3\), and \(i=5\), RLIM wins \(62\), \(65\), and \(61\) of the \(77\) comparable operating points, respectively, while the order-\(1\) comparison is also slightly favorable to RLIM, with \(41\) wins against \(36\). The order-\(4\) comparison is more mixed: RLL$_4$ has more pointwise wins, whereas the operating points won by RLIM$_4$ exhibit the larger mean improvement. These results show that the benefit of minimum-weight selection is order-dependent rather than uniform across all operating points.

Table~\ref{tab:overall_winner} summarizes the methods that attain the smallest BER over the \(77\) operating points. In Table~\ref{tab:overall_winner}, the mean dominance ratio is computed only over the operating points where the method is the unique winner; at each such point, it is the BER of the second-best method divided by the BER of the winning method, averaged over those points. RLIM$_3$ has the largest number of best results, with \(37\) unique wins. RLIM$_4$ and RLIM$_5$ follow with \(10\) unique wins each. Uncoded transmission wins in \(8\) operating points. Overall, the strongest performance across the tested set is obtained by moderate-order RLIM codes, in agreement with the panel-level behavior in Fig.~\ref{fig:all_16_shared_legend_3} and with the RLIM-vs.-RLL comparison in Table~\ref{tab:rlim-vs-rll-elementwise}.

\section{Conclusion}
\label{sec:conclusion}

This paper developed an enumerative realization of weight-minimizing RLIM codes for diffusion-based molecular communication. The original RLIM construction was already one of the most promising constrained-coding approaches for MC, since it combines run-length-limited ISI suppression with minimum-weight codebook selection. This paper shows that, among the prominent MC coding methods compared here, moderate-order RLIM codes attain the strongest overall BER performance across the tested operating points. However, the original realization required storing the entire size-\(2^k\) RLIM codebook, with a memory cost of \(n\cdot 2^k\) bits. This made RLIM difficult to use beyond moderate information dimensions and unsuitable as a realistic implementation model for resource-constrained MC transmitters and receivers.

The proposed realization removes this full-codebook storage barrier without changing the underlying RLIM code design. By using constant-weight \((i,\infty)\)-RLL enumeration, Cover-style ranking and unranking, and cumulative weight-layer offsets, the encoder and decoder recover the same minimum-weight RLIM ordering used by the original construction. The selected codebooks and the projection-based decoding convention are therefore preserved, while the original non-enhanced realization requires only the rank shift described in Section~\ref{sec:algorithms}; the codewords themselves no longer need to be stored. The resulting representation uses only polynomial-size counting tables, replacing the exponentially growing full-codebook representation.

This change has two consequences. First, it makes RLIM much more plausible as an implementable coding method for future resource-constrained MC systems, where memory and preprocessing limitations are central. Second, it unlocks larger information dimensions that were difficult to explore under the full-codebook realization. The reported \(k\)-sweeps show that information dimensions beyond the \(k=16\) regime considered with the original full-codebook realization can improve the BER performance of individual RLIM orders, indicating that the previous storage bottleneck prevented access to potentially advantageous larger-dimensional operating points. Overall, the proposed realization turns RLIM from a storage-limited codebook construction into a scalable and practically usable constrained-coding method, while preserving the performance advantages that make it one of the strongest tested coding approaches for MC.

\section*{Acknowledgment}

This work is dedicated to the memory of Muzaffer \c{S}ahin, the first author's late grandfather.

\bibliographystyle{IEEEtran}
\bibliography{References}

@article{Akyildiz2010,
author = {Akyildiz, Ian F. and Jornet, Josep Miquel},
title = {The internet of nano-things},
year = {2010},
issue_date = {December 2010},
publisher = {IEEE Press},
volume = {17},
number = {6},
issn = {1536-1284},
url = {https://doi.org/10.1109/MWC.2010.5675779},
doi = {10.1109/MWC.2010.5675779},
journal = {Wireless Commun.},
month = dec,
pages = {58–63},
numpages = {6}
}

@article{Cover1973,
  author  = {T. M. Cover},
  title   = {Enumerative Source Encoding},
  journal = {IEEE Transactions on Information Theory},
  volume  = {19},
  number  = {1},
  pages   = {73--77},
  month   = jan,
  year    = {1973},
  doi     = {10.1109/TIT.1973.1054929}
}

@ARTICLE{BeenkerImmink1983,
  author={Beenker, G. and Immink, K.},
  journal={IEEE Transactions on Information Theory}, 
  title={A generalized method for encoding and decoding run-length-limited binary sequences (Corresp.)}, 
  year={1983},
  volume={29},
  number={5},
  pages={751-754},
  doi={10.1109/TIT.1983.1056728}}

@article{Kurmaev2002,
  author  = {O. F. Kurmaev},
  title   = {Enumerative coding for constant-weight binary sequences with constrained run-length of zeros},
  journal = {Problems of Information Transmission},
  volume  = {38},
  number  = {4},
  pages   = {249--254},
  year    = {2002}
}

@ARTICLE{Kurmaev2009,
  author={Kurmaev, Oleg F.},
  journal={IEEE Transactions on Information Theory}, 
  title={Constant-Weight and Constant-Charge Binary Run-Length Limited Codes}, 
  year={2011},
  volume={57},
  number={7},
  pages={4497-4515},
  doi={10.1109/TIT.2011.2145490}}

@article{HareedyDabakCalderbank2022,
author = {Hareedy, Ahmed and Dabak, Beyza and Calderbank, Robert},
title = {The Secret Arithmetic of Patterns: A General Method for Designing Constrained Codes Based on Lexicographic Indexing},
year = {2022},
issue_date = {Sept. 2022},
publisher = {IEEE Press},
volume = {68},
number = {9},
issn = {0018-9448},
url = {https://doi.org/10.1109/TIT.2022.3170692},
doi = {10.1109/TIT.2022.3170692},
journal = {IEEE Trans. Inf. Theor.},
month = sep,
pages = {5747–5778},
numpages = {32}
}

@ARTICLE{SahinAkan2024RLIM,
  author={Şahin, Melih and Akan, Ozgur B.},
  journal={IEEE Transactions on Molecular, Biological, and Multi-Scale Communications}, 
  title={Run-Length-Limited ISI-Mitigation (RLIM) Coding for Molecular Communication}, 
  year={2026},
  volume={12},
  number={},
  pages={376-387},
  doi={10.1109/TMBMC.2026.3659828}}

@book{Nakano2013, place={Cambridge}, title={Molecular Communication}, DOI={10.1017/CBO9781139149693}, publisher={Cambridge University Press}, author={Nakano, Tadashi and Eckford, Andrew W. and Haraguchi, Tokuko}, year={2013}}

@ARTICLE{Kuscu2019,
  author={Kuscu, Murat and Dinc, Ergin and Bilgin, Bilgesu A. and Ramezani, Hamideh and Akan, Ozgur B.},
  journal={Proceedings of the IEEE}, 
  title={Transmitter and Receiver Architectures for Molecular Communications: A Survey on Physical Design With Modulation, Coding, and Detection Techniques}, 
  year={2019},
  volume={107},
  number={7},
  pages={1302-1341},
  doi={10.1109/JPROC.2019.2916081}}

@ARTICLE{Akan2017Fund,
  author={Akan, Ozgur B. and Ramezani, Hamideh and Khan, Tooba and Abbasi, Naveed A. and Kuscu, Murat},
  journal={Proceedings of the IEEE}, 
  title={Fundamentals of Molecular Information and Communication Science}, 
  year={2017},
  volume={105},
  number={2},
  pages={306-318},
  doi={10.1109/JPROC.2016.2537306}}

@ARTICLE{Kislal2020,
  author={Kislal, A. Oguz and Akdeniz, Bayram Cevdet and Lee, Changmin and Pusane, Ali E. and Tugcu, Tuna and Chae, Chan-Byoung},
  journal={IEEE Access}, 
  title={ISI-Mitigating Channel Codes for Molecular Communication Via Diffusion}, 
  year={2020},
  volume={8},
  number={},
  pages={24588-24599},
  doi={10.1109/ACCESS.2020.2970108}}

@ARTICLE{Shih2013,
  author={Shih, Po-Jen and Lee, Chia-Han and Yeh, Ping-Cheng and Chen, Kwang-Cheng},
  journal={IEEE Journal on Selected Areas in Communications}, 
  title={Channel Codes for Reliability Enhancement in Molecular Communication}, 
  year={2013},
  volume={31},
  number={12},
  pages={857-867},
  doi={10.1109/JSAC.2013.SUP2.12130018}}

@article{Hamming1950,
author = {Hamming, R. W.},
title = {Error Detecting and Error Correcting Codes},
journal = {Bell System Technical Journal},
volume = {29},
number = {2},
pages = {147-160},
doi = {https://doi.org/10.1002/j.1538-7305.1950.tb00463.x},
url = {https://onlinelibrary.wiley.com/doi/abs/10.1002/j.1538-7305.1950.tb00463.x},
eprint = {https://onlinelibrary.wiley.com/doi/pdf/10.1002/j.1538-7305.1950.tb00463.x},
year = {1950}
}

@article{BaiLeeson2014,
author = {Bai, Chenyao and Leeson, Mark S. and Higgins, Matthew David},
title = {Minimum energy channel codes for molecular communications},
journal = {Electronics Letters},
volume = {50},
number = {23},
pages = {1669-1671},
doi = {https://doi.org/10.1049/el.2014.3345},
url = {https://ietresearch.onlinelibrary.wiley.com/doi/abs/10.1049/el.2014.3345},
eprint = {https://ietresearch.onlinelibrary.wiley.com/doi/pdf/10.1049/el.2014.3345},
year = {2014}
}

@ARTICLE{channel_characteristics,
  author={Yilmaz, H. Birkan and Heren, Akif Cem and Tugcu, Tuna and Chae, Chan-Byoung},
  journal={IEEE Communications Letters}, 
  title={Three-Dimensional Channel Characteristics for Molecular Communications With an Absorbing Receiver}, 
  year={2014},
  volume={18},
  number={6},
  pages={929-932},
  doi={10.1109/LCOMM.2014.2320917}}

@INPROCEEDINGS{multinomial,
  author={Vakilipoor, Fardad and Barletta, Luca and Bregni, Stefano and Magarini, Maurizio},
  booktitle={2024 IEEE Latin-American Conference on Communications (LATINCOM)}, 
  title={Correlated Source Achievable Information Rate Analysis in Diffusive Channels with Memory}, 
  year={2024},
  volume={},
  number={},
  pages={1-6},
  doi={10.1109/LATINCOM62985.2024.10770672}}

@misc{ldp,
      title={Local Differential Privacy for Molecular Communication Networks}, 
      author={Melih Şahin and Ozgur B. Akan},
      year={2026},
      eprint={2603.00690},
      archivePrefix={arXiv},
      primaryClass={cs.IT},
      url={https://arxiv.org/abs/2603.00690}, 
}

@article{recursion,
title = {Block codes for a class of constrained noiseless channels},
journal = {Information and Control},
volume = {17},
number = {5},
pages = {436-461},
year = {1970},
issn = {0019-9958},
doi = {https://doi.org/10.1016/S0019-9958(70)90369-4},
url = {https://www.sciencedirect.com/science/article/pii/S0019995870903694},
author = {D.T. Tang and L.R. Bahl}
}

@ARTICLE{normalization,
  author={Gursoy, Mustafa Can and Basar, Ertugrul and Pusane, Ali Emre and Tugcu, Tuna},
  journal={IEEE Transactions on Communications}, 
  title={Index Modulation for Molecular Communication via Diffusion Systems}, 
  year={2019},
  volume={67},
  number={5},
  pages={3337-3350},
  doi={10.1109/TCOMM.2019.2898665}}

@ARTICLE{ISI_mitigating_methods_2015,
  author={Tepekule, Burcu and Pusane, Ali E. and Yilmaz, H. Birkan and Chae, Chan-Byoung and Tugcu, Tuna},
  journal={IEEE Transactions on Molecular, Biological and Multi-Scale Communications}, 
  title={ISI Mitigation Techniques in Molecular Communication}, 
  year={2015},
  volume={1},
  number={2},
  pages={202-216},
  doi={10.1109/TMBMC.2015.2501745}}

@article{ReedSolomon1960,
author = {Reed, I. S. and Solomon, G.},
title = {Polynomial Codes Over Certain Finite Fields},
journal = {Journal of the Society for Industrial and Applied Mathematics},
volume = {8},
number = {2},
pages = {300-304},
year = {1960},
doi = {10.1137/0108018},

URL = { 
    
        https://doi.org/10.1137/0108018
    
    

},
eprint = { 
    
        https://doi.org/10.1137/0108018
    
    

}

}

@ARTICLE{Dissanayake2017RS,
  author={Dissanayake, Maheshi B. and Deng, Yansha and Nallanathan, Arumugam and Ekanayake, E. M. N. and Elkashlan, Maged},
  journal={IEEE Communications Letters}, 
  title={Reed Solomon Codes for Molecular Communication With a Full Absorption Receiver}, 
  year={2017},
  volume={21},
  number={6},
  pages={1245-1248},
  doi={10.1109/LCOMM.2017.2671858}}

@INPROCEEDINGS{BCSK,
  author={Kuran, M. S. and Yilmaz, H. B. and Tugcu, T. and Akyildiz, I. F.},
  booktitle={2011 IEEE International Conference on Communications (ICC)}, 
  title={Modulation Techniques for Communication via Diffusion in Nanonetworks}, 
  year={2011},
  volume={},
  number={},
  pages={1-5},
  doi={10.1109/icc.2011.5962989}}

\begin{IEEEbiography}
[{\includegraphics[width=1in,height=1.25in,clip,keepaspectratio]{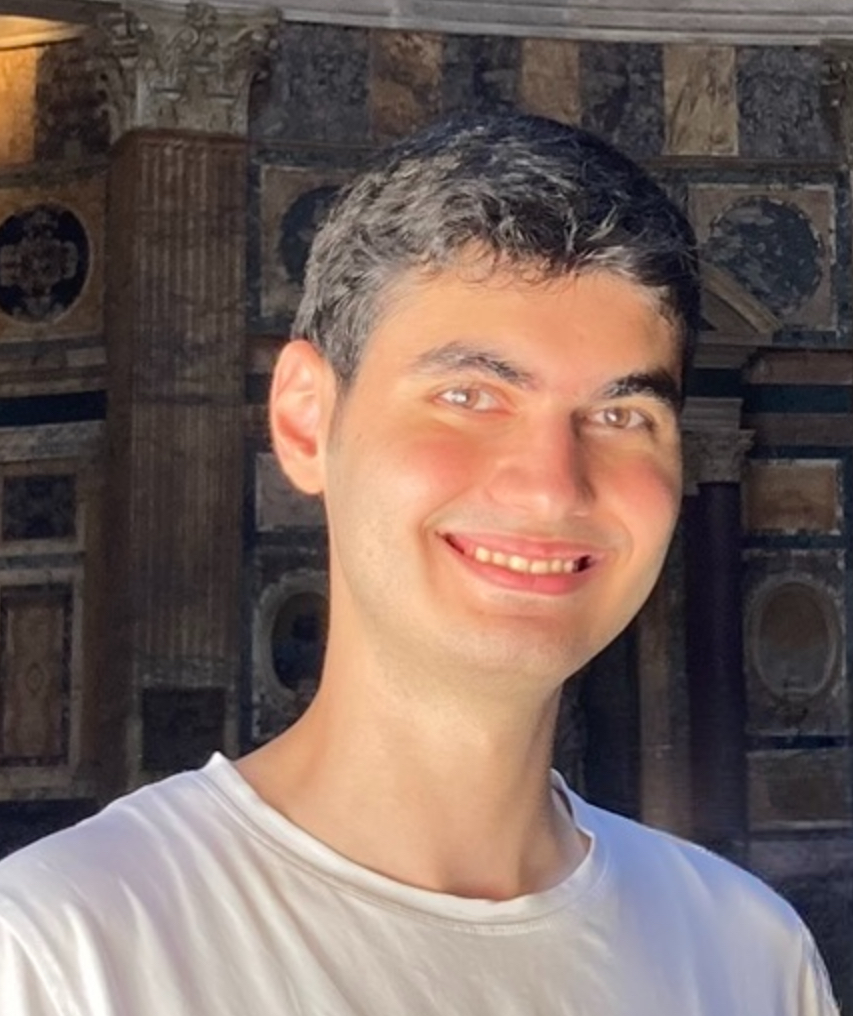}}]{Melih Şahin (Graduate Student Member, IEEE)}
is a PhD student in the Department of Engineering at the University of Cambridge. He received an MS in Electrical and Electronics Engineering with thesis in 2025 and a BSc in Computer Engineering with a Mathematics minor and an AI track in 2024 at Koç University, after his first year of undergraduate studies at KAIST. His notable distinctions include an Intel ISEF 4th place Grand Award in Mathematics in 2018. His research interests cover Information Theory, Coding Theory, and Molecular Communication.
\end{IEEEbiography}

\begin{IEEEbiography}[{\includegraphics[width=1in,height=1.25in,clip,keepaspectratio]{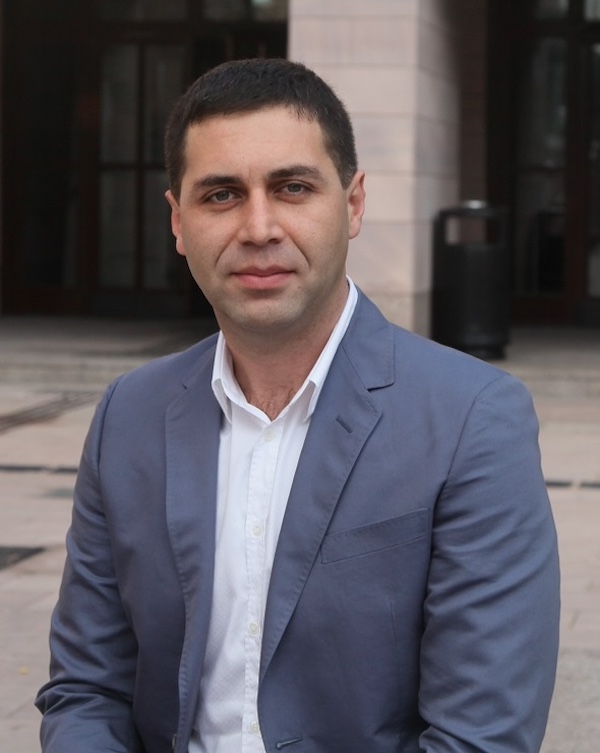}}]{Ozgur B. Akan (Fellow, IEEE)}
received the PhD from the School of Electrical and Computer Engineering, Georgia Institute of Technology, Atlanta, in 2004. He is currently the Head of Centre for neXt Communications (CXC), with the Department of Engineering, University of Cambridge, U.K., and the Director of Centre for neXt Communications (CXC), Koç University, Türkiye. His research interests include wireless, nano, and molecular communications and Internet of Everything.
\end{IEEEbiography}

\end{document}